\documentclass{article}
\usepackage{amsmath}
\usepackage{amsfonts}
\usepackage{amssymb}
\usepackage{graphicx}
\usepackage{float}
\usepackage{booktabs}
\usepackage[top=2cm, bottom=2cm]{geometry}
\usepackage{lineno}

\begin{document}
\title{Confidence, Statistical Evidence and Relative Belief with Applications to a Problem in Particle Physics}
\author{Michael Evans\footnote{mevansthree.evans@utoronto} and Siqi Zheng\footnote{ timothyzheng2000@gmail.com} \\Department of Statistical Sciences, University of Toronto}

\date{ }

\maketitle

\begin{abstract}
Probability theory provides a clear definition of what is meant by evidence in favor, against or none either way, of an event occurring for an unobserved response, via the principle of evidence. This is immediately applicable when carrying out a proper Bayesian analysis. Even without a prior, this imposes restrictions on reported inferences as these need to reflect the likelihood ordering. Relative belief inferences satisfy this requirement and, when the errors in these inferences are controlled, they also satisfy repeated sampling, or frequentist, requirements such as achieving given confidence levels. Relative belief inferences are considered here for the construction of intervals for uncertainty quantification in the context of a Poisson model for a signal with background noise. These intervals are contrasted with the well-known Feldman-Cousins intervals for this problem. 
\end{abstract}


\section{Introduction}

The purpose of a statistical analysis in any scientific domain is to answer
questions of interest in the following sense: observed data contains evidence
concerning the answers and the goal of a statistical analysis is to reveal
what the evidence indicates, and how strong that indication is, as clearly as
possible. For example, we might want to know if there is evidence in favor of
or against the statement that a new elementary particle has been detected in
an experiment and how strong that evidence is.

As uncontroversial as this goal may seem, there has been considerable
discussion about this in the statistical literature for many years and, in
part, this has focused on the difficulties that arise in specifying exactly
what is meant by statistical evidence. For example, going back to the 1930's
there was a debate between two of the most significant figures in the field,
Fisher and Neyman. Fisher took the evidential approach and Neyman
took the behavioristic approach where the goal is to obtain inferences that minimize some
measure of error under counterfactual repeated sampling. Evans (2024) reviews
the various attempts at defining statistical evidence, including a current
proposal as outlined in Section 2.2, and makes the point that is natural to
combine both approaches although they play different roles:

\begin{quote}
The evidential approach is for inference while the behavioristic approach is
for design to ensure the reliability of the inferences drawn.
\end{quote}

The discussion here is in the context of experiments in particle physics where the observations are of counts modeled by a Poisson$(\lambda+b)$ distribution. For a given time interval, $\lambda$ is the unknown mean number of new particles produced and $b$\ is the mean number of background particles and $b$ may be treated as either known or unknown. So, based on the observed data, we want to determine if there is evidence that a new particle has been discovered and the strength of
this evidence. While much of the discussion is restricted here to this problem, the developments discussed apply much more widely.

One approach to analysis in this context has been the use of the
Feldman-Cousins (1998) confidence interval (FCCI) for $\lambda$. This interval is a valid expression of confidence as it provides a lower bound on the frequentist coverage for the true value of $\lambda$ without being overly conservative, at least when $b$ is known. The construction follows the prescription in Neyman (1937) for the construction of confidence intervals. The FCCI also corrects for a problem that many proposed confidence intervals suffer from when the parameter of interest is restricted in some sense. For example, when sampling from a Poisson$(\lambda)$ distribution, an approximate $\gamma$-confidence interval for $\lambda$ based on likelihood
asymptotics, can sometimes contain negative values. This does not make sense as it is known categorically that $\lambda\geq0.$ Typically, the length of a $\gamma$-confidence interval is used as a measure of the accuracy of an estimator contained in it. As such, the simple solution to this problem, by just truncating the interval to exclude negative values, does not represent a correct assessment of the accuracy of the estimate.

The issue raised here is that many proposed confidence regions, including the FCCI, do not represent evidence properly. This claim is based on a natural characterization of evidence that arises within the context of a probability model as discussed in Section 2.2 and referred to here as the \textit{principle of evidence}. This imposes the requirement that, a region that claims to represent, in any sense, the evidence in a statistical context, must be a likelihood region; see Theorem 1. This criticism is not surprising because the concept of confidence itself was not designed to reflect evidence, but rather confidence regions are used to measure the error in an estimate lying within the region. Certainly, the consideration of such error is essential as it is a measure of the reliability of the inference being quoted. As will be shown in Section 2.2, a relative belief region known as the \textit{plausible region}, is always a likelihood region, contains all the values of the parameter of interest that have evidence in their favor and no others, has a posterior probability that represents the belief that it contains the true value, and can be designed to satisfy a prescribed frequentist confidence as well as a prescribed coverage of meaningfully false values. It is our claim that the plausible region is more appropriate for scientific contexts than confidence regions in general.

Relative belief inferences are Bayesian in nature and are restricted to the use of proper priors. This places the discussion within the context of one probability model and so the principle of evidence applies. Objections are sometimes
raised concerning the use priors but, as will be discussed, they are choices made by an analyst just as the sampling model is typically such a choice. The sampling model can be checked against the objective data via model checking procedures and the prior can be checked via checking for prior-data conflict;
see Section 2.3. In fact it is much easier to replace a defective prior than a defective model.

The book Lyons (1986) describes many of the statistical issues confronted in particle physics. While the relative belief approach is applicable to general statistical inference problems, the focus here is on the problem of Poisson counts where there is background noise. There has been much discussion of the problem of forming confidence intervals for the Poisson with background with Feldman and Cousins (1998) being a centrally important paper. Other papers on this include Cousins and Highland
(1992), Roe and Woodroofe (1999, 2000), Mandelkern (2002), Conrad et al. (2003), Biller and Oser (2015) and Lu et al, (2023). Given that relative belief inferences are a form of Bayesianism, the papers Cousins (1995), Biller
and Oser (2015), and the online notes of Oser (see references for link), are also relevant to this discussion. A general discussion of relative belief inference can be found in Evans
(2015).The relative belief approach has been used in statistical analyses in physics previously. For example, Gu et al. (2019), Teo et al. (2024a, b) present applications to problems in quantum mechanics. Also, see Shang et al. (2013) and Englert (2026) where the construction of error regions are the same as relative belief credible regions.

In Section 2, the basic formulation of a statistical problem is presented. In Section 2.1 the construction of FCCI's is discussed
and it is shown that these intervals violate the likelihood ordering. In Section 2.2, the principle of
evidence is introduced as well as the associated relative belief inferences. Section 2.3 is concerned with checking for prior-data conflict and Section 2.4 deals with the repeated sampling behavior of relative belief inferences.
Section 3 discusses the problem of particle detection with both fixed and uncertain background. In Section 4 a reference is provided for access to the code used for the computations conducted in the paper and in Section 5
some conclusions are drawn.

\section{Statistical Analyses}

Consider a standard formulation of a statistical problem. There is a
\textit{statistical model} $\{p(\cdot\,|\,\theta):\theta\in\Theta\},$ namely,
a collection of probability distributions, as represented by their densities
with respect to some common support measure on a set $\mathcal{X}$ (the
\textit{sample space}) and indexed by \textit{model parameter} $\theta
\in\Theta$ (the \textit{parameter space}). The observed data $x\in\mathcal{X}$
is assumed to have been generated from a process that produces data from one
of these probability distributions. The true value of $\theta,$ namely, the
value of the parameter corresponding to the distribution which generated the
data, will be labelled labelled $\theta_{true}.$ The pair $I=(\{p(\cdot
\,|\,\theta):\theta\in\Theta\},x)$ is called an \textit{inference base} for
any theory of statistical reasoning that uses these, and only these,
ingredients to form inferences about some object of interest $\psi=\Psi
(\theta).$ Here $\Psi(\theta)$ is some characteristic of the distribution
$p(\cdot\,|\,\theta),$ and the goal is ultimately to know $\psi_{true}%
=\Psi(\theta_{true}).$ The values $\theta\in\Psi^{-1}\{\psi\}$ are known as
nuisance parameters and it is significant problem to deal with these in
a logically sound way, based only upon the inference base $I.$

Other ingredients are often added to $I$ and, in particular, we shall be
concerned with adding a \textit{prior} probability distribution on $\Theta,$
represented by its density $\pi,$ that expresses beliefs about where the true
value of $\theta$ lies in $\Theta.$ The prior can be factored as $\pi
(\theta)=\pi(\theta\,|\,\psi)\pi_{\Psi}(\psi)$ where $\pi(\cdot\,|\,\psi)$ is
the density of $\theta$ conditioned to $\Psi^{-1}\{\psi\},$ namely, the prior
density on the nuisance parameters, and $\pi_{\Psi}$ is the the prior on the
parameter of interest. The handling of nuisance parameters is conceptually
simple in Bayesian inference as these are just integrated out to produce a new
statistical model $\{m(\cdot|\,\psi:\psi\in\Psi(\Theta)\}$ where
\[
m(x\,|\,\psi)=\int_{\Psi^{-1}\{\psi\}}p(x\,|\,\theta)\pi(\theta\,|\,\psi
)\,d\theta.
\]
The reduction from the Bayesian inference base for inferences about $\theta,$
namely, $I^{Bayes}=(\{p(\cdot\,|\,\theta):\theta\in\Theta\},\pi,x)$ to
$I_{\Psi}^{Bayes}=(\{m(\cdot\,|\,\psi):\psi\in\Psi(\Theta)\},\pi_{\Psi},x)$
for inferences about $\psi,$ is then logically consistent and unambiguous.

There are two basic problems/questions that any theory of statistical
inference needs to answer.

\begin{quote}
$\mathbf{H}$: Is there evidence that the hypothesis $H_{0}:\psi_{true}%
=\psi_{0}$ is true or false and how strong is this evidence?

$\mathbf{E}$: What value $\psi(x)\in\Psi(\Theta)$, among those having evidence
in its favor of being the true value, should be used to estimate $\psi$ and
how accurate is the estimate $\psi(x)?$
\end{quote}

\subsection{Feldman-Cousins Confidence Regions}

One notable attempt to characterize statistical evidence based on inference
base $I$ is provided by Royall (1997) who attempted to do this using the
likelihood function alone without invoking repeated sampling. In the context
of the full model parameter, Royall's characterization was provided in terms
of the relative likelihood function%
\[
L^{rel}(\theta\,|\,x)=\frac{p(x\,|\,\theta)}{\sup_{\theta \in \Omega}p(x\,|\,\theta)}\in
\lbrack0,1]
\]
where, for some constant $c\in\lbrack0,1]$, a likelihood region $C(x)=\{\theta
:L^{rel}(\theta\,|\,x)\geq c\}$ is to be quoted as representing those values
that have evidence in their favor based on the observed data $x.$ There is no
clear choice for $c,$ however, to provide a cut-off for evidence in favor and
against. Royall (1997) gave an argument, based on an urn model, for choices of
$c$ to reflect weak evidence and strong evidence in favor, but there is no
reason to suppose that such a context can serve as a gold standard for all
other applications.

There is implicit in this usage the idea that the scale $[0,1]$ for the
relative likelihood, has a universal interpretation in terms of evidence. This
means that two different contexts that give rise to the same value of the
relative likelihood are supposed to have the same interpretation. The following simple
example shows, however, that there are good reasons to doubt this, even for the same
model but with different data.\smallskip

\newpage
\noindent\textbf{Example 2.1.}

Suppose the model and the relative likelihoods are as given in Table 1 where the
sample space is $\mathcal{X}=\{1,2\}$ and the parameter space is
$\Theta=\{\theta_{1},\theta_{2}\}.$ Looking at the relative likelihoods it would appear that, when $N$
is large and $x=2$ is observed, there is considerable support, as defined by
the relative likelihood, for $\theta_{2}$ over $\theta_{1},$ as $L^{rel}%
(\theta_{2}\,|\,2)/L^{rel}(\theta_{1}\,|\,2)\rightarrow\infty$ as
$N\rightarrow\infty.$ When $x=1$ is observed, then the support for $\theta
_{1}$ over $\theta_{2}$ seems much weaker as $L^{rel}(\theta_{1}%
\,|\,1)/L^{rel}(\theta_{2}\,|\,1)\rightarrow2.$ On the other hand, with
$\theta(x)$ equal to the MLE, the coverage probabilities $P_{\theta_{1}}(\theta
(x)=\theta_{1})=(N-1)/N\approx1$ and $P_{\theta_{2}}(\theta(x)=\theta
_{2})=1/2$ suggest the opposite. $\blacksquare\smallskip$

%TCIMACRO{\TeXButton{B}{\begin{table}[H] \centering}}%
%BeginExpansion
\begin{table}[t] \centering
%EndExpansion%
\begin{tabular}
[c]{c|cccc}
& $p(1\,|\,\theta)$ & $L^{rel}(\cdot\,|\,1)$ & $p(2\,|\,\theta)$ &
$L^{rel}(\cdot\,|\,2)$\\\hline
$\theta_{1}$ & $(N-1)/N$ & $1$ & $1/N$ & $2/N$\\
$\theta_{2}$ & $1/2$ & $N/2(N-1)$ & $1/2$ & $1$%
\end{tabular}
\caption{Model and relative likelihoods for Example 2.1.}\label{TableKey copy(4)}%
%TCIMACRO{\TeXButton{E}{\end{table}}}%
%BeginExpansion
\end{table}%
%EndExpansion

There are many reasons to doubt that the relative likelihood is the complete
characterization of statistical evidence. In particular, the need to rely on the profile likelihood for $\psi$,
$$
L_\Psi(\psi \,|\,x) =  \sup_{\theta \in \Psi^{-1} \{\psi\}} p(x \,|\, \theta),
$$ 
for inferences about a marginal parameter $\psi$, is a significant weakness. The profile
likelihood is, in general, not a likelihood, namely, it is not proportional to the probability of the observed data based on some model. By contrast, after integrating out
nuisance parameters, the inference base $I_{\Psi}=(\{m(\cdot\,|\,\psi):\psi
\in\Psi(\Theta)\},x)$ gives rise to a valid likelihood for $\psi.$ This is important because the fundamental idea that underlies the use of the likelihood is the total ordering of the values of $\psi$ induced by the likelihood. In particular,
$\psi_{1}$ is preferred to $\psi_{2}$ whenever $m(x|\,\psi_{1})>m(x|\,\psi
_{2})$ because the observed data is more probable when $\psi_{1}$ is the true
value than when $\psi_{2}$ is the true value. The likelihood ordering seems
very natural and Theorem 1 in Section 2.2 implies that any region quoted as a
candidate to contain the true value, must respect the ordering and so be a
likelihood region. To use the likelihood as a basis for inference and then
discard this ordering, is contradictory as it violates the very foundational
concept that motivated the approach.

Feldman and Cousins (1998) use a methodology for constructing a  confidence interval based on Neyman (1937). The Feldman-Cousins technique is a novel and intuitive approach based on the relative
likelihood. This can be applied quite generally, at least provided there are no nuisance parameters. For this consider acceptance region $A(\theta,c)=\{x:L^{rel}%
(\theta\,|\,x)\geq c\}$ for each $\theta\in\Theta$, with the interpretation
that the hypothesis $H_{0}:\theta_{true}=\theta$ is accepted as true whenever
$x\in A(\theta,c)$ and rejected otherwise. Then for a given $\theta$ and confidence $\gamma\in(0,1),$ the
value $c$ is determined as
\[
c_{\gamma}(\theta)=\sup\{c:P_{\theta}(A(\theta,c))\geq\gamma\}.
\]
The $\gamma$-confidence region is then given by $C(x)=\{\theta:x\in
A(\theta,c_{\gamma}(\theta))\}$ as $P_{\theta}(\theta\in C(x))=P_{\theta
}(A(\theta,c_{\gamma}(\theta)))\geq\gamma$ for every $\theta\in\Theta.$ Note that this construction implicitly suggests that accepting $H_{0}:\theta_{true}=\theta$ means that there is evidence in favor of $\theta$, but this is dependent on $\gamma$ which is arbitrary in the sense that it isn't clear what value it should be for this role.

The Feldman-Cousins construction is illustrated in the following example. This example also shows that the
technique does not generally produce a likelihood region and can result in what is known as an improper confidence region. 
\smallskip

\noindent\textbf{Example 2.2.} 

Suppose the model and relative likelihoods are as given in Table 2 where
$\mathcal{X}=\{1,2,3\}$ and $\Theta=\{\theta_{1},\theta_{2},\theta_{3}\}$.%
The Feldman-Cousins construction leads to the following possible acceptance regions for
$\theta,$ together with their probability contents

%BeginExpansion
\begin{table}[tb] \centering
%EndExpansion%
\begin{tabular}
[c]{c|cccccc}
& $p(1\,|\,\theta)$ & $L^{rel}(\theta\,|\,1)$ & $p(2\,|\,\theta)$ &
$L^{rel}(\theta\,|\,2)$ & $p(3\,|\,\theta)$ & $L^{rel}(\theta\,|\,3)$\\\hline
$\theta_{1}$ & $5/12$ & $1$ & $6/12$ & $6/8$ & $1/12$ & $1/5$\\
$\theta_{2}$ & $3/12$ & $3/5$ & $4/12$ & $4/8$ & $5/12$ & $1$\\
$\theta_{3}$ & $4/12$ & $4/5$ & $8/12$ & $1$ & $0$ & $0$%
\end{tabular}
\caption{Model and relative likelihoods for Example 2.2.}\label{TableKey copy(2)}%
%TCIMACRO{\TeXButton{E}{\end{table}}}%
%BeginExpansion
\end{table}%
%EndExpansion

\noindent %
\begin{align*}
&A\left(\theta_{1},\frac{1}{5}\right)=\mathcal{X},P_{\theta_1}(\mathcal{X})=1,
A\left(\theta_{1},\frac{6}{8}\right)=\{1,2\},P_{\theta_{1}}(\{1,2\})=\frac{11}{12},
A\left(\theta_{1},1\right)=\{1\},P_{\theta_{1}}(\{1\})=\frac{5}{12}
\\
&A\left(\theta_{2},\frac{4}{8}\right)=\mathcal{X},P_{\theta_2}(\mathcal{X})=1,
A\left(\theta_{2},\frac{3}{5}\right)=\{1,3\},P_{\theta_{2}}(\{1,3\})=\frac{8}{12},
A\left(\theta_{2},1\right)=\{3\},P_{\theta_{2}}(\{3\})=\frac{5}{12}
\\
&A\left(\theta_{3},0\right)=\mathcal{X},P_{\theta_3}(\mathcal{X})=1,
A\left(\theta_{3},\frac{4}{5}\right)=\{1,2\},P_{\theta_{3}}(\{1,2\})=1,
A\left(\theta_{3},1\right)=\{2\},P_{\theta_{3}}(\{2\})=\frac{8}{12}.
\end{align*}

From these values, and with confidence lower bound $\gamma=1/2$,
\begin{align*}
&c_{1/2}(\theta_{1})=\frac{6}{8}, A\left(\theta_{1},\frac{6}{8}\right)=\{1,2\},
\\
&c_{1/2}(\theta_{2})=\frac{3}{5},A\left(\theta_{2},\frac{3}{5}\right)=\{1,3\},
\\
&c_{1/2}(\theta_{3})=1,A\left(\theta_{3},1\right)=\{2\}.
\end{align*}
This leads to the $1/2$-confidence region
$C(1)=\{\theta_{1},\theta_{2}\},C(2)=\{\theta_{1},\theta_{3}\},C(3)=\{\theta
_{2}\}$
 with coverages
 $$
P_{\theta_{1}}(\theta_{1}\in C(x))=11/12,P_{\theta_{2}}(\theta_{2}\in
C(x))=8/12,P_{\theta_{3}}(\theta_{3}\in C(x))=8/12.
$$
Therefore, the actual confidence lower bound, is $2/3.$ 
Notice, however that $C(1)=\{\theta_1,\theta_2\}$ is not a likelihood region because $p_{\theta_{3}}(1)>p_{\theta_{2}}(1)$ and
$\theta_{3}\notin C(1)$ while $\theta_{2}\in C(1).$ 
If $\theta_{3}$ was just added
to form the likelihood region $C(1)=\{\theta_{1},\theta_{2},\theta_{3}\},$ then, as discussed below, this is not appropriate. 

Now consider the case where $\gamma=11/12$. This leads to 
\begin{align*}
&c_{11/12}(\theta_{1})=6/8, A\left(\theta_{1},11/12\right)=\{1,2\},
\\
&c_{11/12}(\theta_{2})=4/8,A\left(\theta_{2},4/8\right)=\{1,2,3\},
\\
&c_{11/12}(\theta_{3})=4/5,A\left(\theta_{3},4/5\right)=\{1,2\},
\end{align*}
and the $11/12$-confidence region 
$C(1)=\Theta,C(2)=\Theta,
C(3)=\{\theta_{2}\}$
with coverage probabilities
$$
P_{\theta_{1}}(\theta_{1}\in C(x))=11/12,P_{\theta_{2}}(\theta_{2}\in
C(x))=1,P_{\theta_{3}}(\theta_{3}\in C(x))=1.
$$
Notice that $C(1) = C(2)= \Theta$, the whole parameter space. Such an outcome is commonly considered as an absurdity and a region that exhibits such behavior is called \textit{improper} or \textit{absurd}. The reason for this is that, if $C$(1) or $C$(2) is stated, then it is categorically known that the true value is in the set and the confidence level $11/12$ seems irrelevant. It has been proven in Evans et al. (2024), that the regions that arise via relative belief can
never be equal to the whole parameter space. In particular, when a region equals $\Theta$, the region necessarily contains values for which there is evidence against them being the true value. Furthermore, it is not the case that such an outcome means that no evidence concerning the true value has been obtained. As discussed in Example 2.5, saying that no evidence has been obtained requires that the likelihood be flat at such a data value, and then the reported region must equal the null set otherwise coverages are being artificially overstated.
$\blacksquare$\smallskip

The fact that the Feldman-Cousins regions do not respect the likelihood ordering, is not just a feature of the discrete
case. For example, consider Table X in Feldman and Cousins (1998), which
provides FCCI's for the mean of a normal with known standard deviation
$\sigma=1.$ Then any interval where $\bar{x}>0,$ the lower bound is greater
than 0, and it is not symmetric about its center, cannot be a likelihood
interval. When $\bar{x}=1.70,\,$then the 95\% interval is $\left(0.06,3.66\right)$ and $\left(0.06+3.66\right)  /2=1.86\neq1.70$ so this is
not a likelihood interval.

Perhaps the primary motivation for the FCCI is that it handles situations where the parameter space is constrained. The relative belief
regions accommodate this requirement without difficulty as well as satisfying the repeated sampling confidence requirement. The pursuit of regions that only satisfy some confidence constraint, can be seen generally to lead to regions that exhibit inappropriate characteristics; see Plante (2020, 2026). 

\subsection{Relative Belief Inferences}
The basic, underlying motivation for relative belief inference comes from a pure probabilistic context where there is a single known probability measure. So, consider a probability model $(\Omega, \mathcal{A},P)$ where $\Omega$ is a sample space, $\mathcal{A}$ is a sigma algebra of subsets of $\Omega$ and $P$ is a probability measure on $\mathcal{A}$. Suppose $\omega \in \Omega$ has occurred, but is not observed, and interest is in whether or not $\omega \in A$ where $A \in \mathcal{A}$ and $P(A)>0.$ Suppose further that the information is provided that $\omega \in C$ where $C \in \mathcal{A}$ where $P(C)>0$. This information leads to belief about $A$ being true changing from $P(A)$ to $P(A\, |\, C)$, the conditional probability of $A$ given $C$. The following principle then characterizes evidence.

\begin{quote}
\textit{Principle of Evidence}: For probability model 
$(\Omega, \mathcal{A},P)$, there is evidence in favor of $A$ being true if $P(A\, |\, C) > P(A)$, there is
evidence against if $P(A\, |\, C) < P(A)$ and no evidence
either way if $P(A\, |\, C)= P(A)$.
\end{quote}

\noindent The basic
idea underlying the principle of evidence has long been considered as an
obvious characterization of statistical evidence and underlies a whole branch
of the philosophy of science known as confirmation theory, see Popper (1968) and Salmon (1973).

The way the ``data'', namely, $C$ is observed to be true, cause belief to change determines whether there us evidence in favor or against $A$ being true. Note that beliefs do not change iff $A$ and $C$ are statistically independent, and then the data tells us nothing about $A$. The change in belief tells us nothing about the strength of the evidence, but a natural way to measure this is to consider the probability  $P(A\, |\, C)$ as this indicates how strongly it is believed what the evidence says. So, if there is evidence in favor of $A$ and   $P(A\, |\, C)$ is large, then there is strong evidence in favor of $A$ and, if there is evidence against $A$ and  $P(A\, |\, C)$ is small, then there is strong evidence against $A$.

In many applications it is important to have an ordering principle based upon the evidence. For example, suppose there is a partition $\{A_1,A_2,\ldots\}$ of $\Omega$ such that $P(A_i)>0$ for $i=1,2,\ldots$ and interest is in which partition element is true.  A natural ordering principle is to use the \textit{relative belief ratio} 
$$
RB(A\,|\,C)=\frac{ P(A\, |\, C)}{P(A)}.
$$
Note that the relative belief ratio is a \textit{valid} measure of evidence because it has an unambiguous cut-off implied by the principle of evidence, namely, $RB(A\,|\,C)> (<) 1$ iff there is evidence in favor (against) $A$ being true. So, if $1<RB(A_i\,|\,C)<RB(A_j\,|\,C)$ then, from the point of view of evidence in favor, $A_i$ has less evidence in its favor than that of $A_j$, etc. 

A natural estimate of the true partition element is then 
$$
A_{i(C)}= \arg\sup_{j\in \{1,2,\ldots\}} RB(A_j\,|\,C).
$$
The accuracy of this estimate can be measured by the size of the \textit{plausible region} 
$$
Pl(C)=\{A_i : RB(A_i\,|\,C)>1\},
$$
and its conditional probability $P(Pl(C)\, |\, C)$. Note that the plausible region is the set of values that have evidence in their favor. 

To assess whether or not $A_{i_0}$ is the true partition element, the relative belief ratio $RB(A_{i_0}\,|\,C)$ tells us immediately whether there is evidence in favor or against this. The strength of this evidence can then be assessed by $P(A_{i_0}\, |\, C)$ but, when the number of elements in the partition is large and each $P(A_i)$ is small, then a better measure is the \textit{strength of the evidence}
\begin{equation}
    Str(A_{i_0}\, |\, C) = P(RB(A_i\,|\,C) \leq RB(A_{i_0}\,|\,C)\, |\, C)
    \label{eq: strength}
\end{equation}
So, when there is evidence in favor of $A_{i_0}$ and (\ref{eq: strength}) is large, there is strong evidence in favor of $A_{i_0}$ as there is little belief that the true partition element has more evidence in its favor. Similarly, when there is evidence against $A_{i_0}$ and (\ref{eq: strength}) is small, there is strong evidence against $A_{i_0}$ as there is little belief that the true partition element has less evidence against it.

All of this can now be translated to the statistical context given by the inference base 
$I_{\Psi}^{Bayes}=(m(\cdot\,|\,\psi):\psi\in\Psi(\Theta)\},\pi_{\Psi},x)$.
This is immediate when the prior $\pi_{\Psi}$
is discrete as now $P$ corresponds to the joint distribution of $(\psi,x)$ and $C=\Psi(\Theta) \times \{x\}$. The relative belief ratio at $\psi$ is then
\begin{equation}
RB_{\Psi}(\psi\,|\,x)=\frac{\pi_{\Psi}(\psi\,|\,x)}{\pi_{\Psi}(\psi)}%
=\frac{m(x\,|\,\psi)}{m(x)}, \label{rb1}%
\end{equation}
where $\pi_{\Psi}(\psi\,|\,x)$ is the posterior density of $\psi$ and $m(x)$ is the
prior density of the data. The second equality in (\ref{rb1}) is immediate because 
$\pi_{\Psi}(\psi\,|\,x)=m(x\,|\,\psi)\pi_{\Psi}(\psi)/m(x)$. When $\pi_{\Psi}$ is
continuous, the relative belief ratio takes the same form as (\ref{rb1}) since, when $N_{\delta}(\psi)$
is a sequence of neighborhoods converging nicely to $\{\psi\}$, and $\Pi_{\Psi
},\Pi_{\Psi}(\cdot\,|\,x)$ denote the prior and posterior probability
measures, then
\begin{equation}
\lim_{\delta\rightarrow0}RB_{\Psi}(N_{\delta}(\psi)\,|\,x)=\lim_{\delta
\rightarrow0}\frac{\Pi_{\Psi}(N_{\delta}(\psi)\,|\,x)}{\Pi_{\Psi}(N_{\delta
}(\psi))}=\frac{\pi_{\Psi}(\psi\,|\,x)}{\pi_{\Psi}(\psi)}=\frac{m(x\,|\,\psi
)}{m(x)} \label{rb11}%
\end{equation}
whenever $\pi_{\Psi}$ is continuous and positive at $\psi.$ 
 
The relative
belief ratio is not the only valid measure of evidence but it has many
excellent properties. For example, it is apparent from the second version in
(\ref{rb1}), that $RB_{\Psi}(\cdot\,|\,x)$ is basically the likelihood
function with respect to the model $\{m(x\cdot|\,\psi):\psi\in\Psi(\Theta)\}.$
With the cut-off of 1 specified to determine evidence in favor or against. There isn't the freedom to multiply the relative belief ratio by
an arbitrary constant or divide by the maximum, as that would destroy its interpretation as a measure of
evidence. Another
valid measure of evidence is given by the Bayes factor, but there are many
reasons to prefer the relative belief ratio, see Al-Labadi, Alzaatreh. and Evans (2025). Note that, just as with likelihood-based inferences, all inferences derived via the relative belief ratio are invariant under
smooth reparameterizations, as any Jacobian appears in the numerator and denominator of (\ref{rb1}) and so cancels. Also, since the asymptotics for likelihood inferences depend on
derivatives of the log-likelihood with respect to the parameter, at least when suitable conditions apply, the frequentist
asymptotics of relative belief inferences are the same.

For the \textbf{H} problem there is an immediate indication of whether there
is evidence in favor of or against $H_{0}:\Psi(\theta)=\psi_{0}$ via the value
$RB_{\Psi}(\psi_{0}\,|\,x).$ The strength of the evidence is then measured by $\pi_{\Psi}(\psi\,|\,x)$ or, when the cardinality of $\Psi(\Theta)$ is large, by 
\begin{equation}
Str_\Psi(\psi_0\,|\,x)=\Pi_{\Psi}(RB_{\Psi}(\psi\,|\,x)\leq RB_{\Psi}(\psi_{0}\,|\,x)\,|\,x),
\label{rb2}%
\end{equation}
where $\Pi_\Psi (\cdot \, |\, x)$ refers to the posterior probability measure of $\psi$, and
with the interpretation of (\ref{rb2}) just as with (\ref{eq: strength}). Upper and lower bounds on $Str_\Psi(\psi_0\,|\,x)$ are obtained in Al-Labadi et al. (2025) (Proposition 9) that can avoid need for the full calculation. For example, $RB_{\Psi}(\psi_{0}\,|\,x)\,|\,x)$ is always an upper bound so, when this is small, there is strong evidence against $\psi_0$. 

For the \textbf{E} problem, the relative belief ratio totally orders
the possible values of $\psi\in\Psi(\Theta)$ as, when $1<RB_{\Psi}(\psi
_{1}\,|\,x)<RB_{\Psi}(\psi_{2}\,|\,x),$ there is more evidence in favor of
$\psi_{2}$ than in favor of $\psi_{1}$ and, when $RB_{\Psi}(\psi
_{1}\,|\,x)<RB_{\Psi\text{ }}(\psi_{2}\,|\,x)<1,$ there is more evidence
against $\psi_{1}$ than against $\psi_{2}.$ This leads to a very natural
estimate of $\psi,$ namely, the relative belief estimate,%
\[
\psi(x)=\arg\sup RB_{\Psi}(\psi\,|\,x)=\arg\sup\frac{\pi_{\Psi}(\psi
\,|\,x)}{\pi_{\Psi}(\psi)}=\arg\sup\frac{m(x\,|\,\psi)}{m(x)},
\]
which is the MLE\ under the model $\{m(\cdot\,|\,\psi):\psi\in\Psi(\Theta)\}.$
This satisfies $RB_{\Psi}(\psi(x)\,|\,x)>1$ and so has the most evidence in
its favor, which is an alternative argument in favor of the MLE as the best
estimate. To assess the accuracy of this estimate there is the
\textit{plausible region,}%
\[
{Pl}_{\Psi}(x)=\{\psi:RB_{\Psi}(\psi\,|\,x)>1\}=\{\psi:m(x\,|\,\psi)>m(x)\},
\]
the set of values of $\psi$ having evidence in their favor. Note that the
second expression in (\ref{rb1}) establishes that ${Pl}_{\Psi}(x)$ is a likelihood region
with respect to the model $\{m(\cdot\,|\,\psi):\psi\in\Psi(\Theta)\}$ and
there is now a clear specification of when a likelihood region displays
evidence in favor. The size of ${Pl}_{\Psi}(x)$ is a measure of the accuracy
of $\psi(x)$ and its posterior content $\Pi_{\Psi}(Pl_{\Psi}(x)\,|\,x)$
measures how strongly to believe what the evidence says. The size measure of
${Pl}_{\Psi}(x)$ could be volume or the prior content $\Pi_{\Psi}({Pl}_{\Psi
}(x))$ or something else that was more relevant to a particular application. It is also possible to quote a $\gamma$-relative belief credible
region $C_{\Psi,\gamma}(x)=\{\psi:RB_{\Psi}(\psi\,|\,x)\geq c_{\gamma}(x)\}$
for $\psi,$ where $c_{\gamma}(x)=\inf\{c:$ $\Pi_{\Psi}(RB_{\Psi}%
(\psi\,|\,x)\leq c\,|\,x)\geq1-\gamma\},$ provided $\gamma\geq\Pi_{\Psi
}(Pl_{\Psi}(x)\,|\,x)$ as otherwise $C_{\Psi,\gamma}(x)$ would contain a value
of $\psi$ for which there is evidence against $\psi$ being the true value.

Now consider some of the previously discussed examples from the
point-of-view of relative belief.\smallskip

\newpage
\noindent\textbf{Example 2.3. }(\textit{Example 2.1 continued})

Suppose a uniform prior $\pi$ is placed on $\Theta=\{\theta_{1},\theta_{2}\}$
and $\Psi(\theta)=\theta.$ The relative belief ratio is given by
\[
RB(\theta\,|\,x)=\frac{p(x\,|\,\theta)}{m(x)}=\left\{
\begin{array}
[c]{cc}%
4(N-1)/(3N-2)\rightarrow4/3, & x=1,\theta=\theta_{1},\\
2N/(3N-2)\rightarrow2/3 & x=1,\theta=\theta_{2},\\
4/(N+2)\rightarrow0 & x=2,\theta=\theta_{1},\\
2N/(N+2)\rightarrow2 & x=2,\theta=\theta_{2},
\end{array}
\right.
\]
where the limiting values as $N \rightarrow \infty$ are also provided.
Provided $N>2,$ then $RB(\theta\,|\,1)>1$ iff $\theta=\theta_{1}$ so
${Pl}_{\Psi}(1)=\left\{  \theta_{1}\right\}  ,$ while $RB(\theta\,|\,2)>1$ iff
$\theta=\theta_{2}$ so ${Pl}_{\Psi}(2)=\left\{  \theta_{2}\right\}  .$ The
posterior contents of the plausible regions are then
\begin{align*}
\Pi({Pl}_{\Psi}(1)\,|\,1)  &  =2(N-1)/(3N-2)\rightarrow2/3,\\
\Pi({Pl}_{\Psi}(2)\,|\,2)  &  =N/(N+2)\rightarrow1.
\end{align*}
Whereas the relative likelihood comparison suggested vastly more support for
the MLE when $x=2,$ than when $x=1,$ calibrating the implications of the
relative belief ratio via the posterior probabilities, shows that this is an
exaggeration. $\blacksquare$\smallskip

Consider now applying relative belief to Example 2.2.\smallskip

\noindent\textbf{Example 2.4 }(\textit{Example 2.2 continued})

Suppose a uniform prior is placed on $\Theta=\{\theta_{1},\theta_{2}%
,\theta_{3}\}.$ The posterior and relative belief ratios are given in Table
2. The plausible regions for $\theta$ and their posterior contents are given
by ${Pl}(1)=\{\theta_{1}\},\Pi({Pl}(1)\,|\,1)=5/12,{Pl}(2)=\{\theta_{3}%
\},\Pi({Pl}(2)\,|\,2)=4/9$ and ${Pl}(3)=\{\theta_{2}\},\Pi({Pl}(3)\,|\,3)=5/6$
and each is a likelihood region. Also, no elements can be added to these
regions without including values for which there is evidence against them
being the true value.%

%TCIMACRO{\TeXButton{B}{\begin{table}[H] \centering}}%
%BeginExpansion
\begin{table}[h] \centering
%EndExpansion%
\begin{tabular}
[c]{c|cccccc}
& $\pi(\theta\,|\,1)$ & $RB(\theta\,|\,1)$ & $\pi(\theta
\,|\,2)$ & $RB(\theta
\,|\,2)$ & $\pi(\theta\,|\,3)$ & $RB(\theta\,|\,3)$\\\hline
$\theta_{1}$ & $5/12$ & $5/4$ & $6/18$ & $1$ & $1/6$ & $1/2$\\
$\theta_{2}$ & $3/12$ & $3/4$ & $4/18$ & $4/6$ & $5/6$ & $5/2$\\
$\theta_{3}$ & $4/12$ & $1$ & $8/18$ & $8/6$ & $0$ & $0$%
\end{tabular}
\caption{Posterior probabilities and relative belief ratios for the model in Example 2.4 when using a uniform prior.}\label{TableKey copy(5)}%
%TCIMACRO{\TeXButton{E}{\end{table}}}%
%BeginExpansion
\end{table}%
%EndExpansion

Of some interest are the values $RB(\theta_{3}\,|\,1)=RB(\theta_{1}\,|\,2)=1$
which indicates that no evidence has been found either way for these values of
$\theta$ when that data has been observed. This implies that, in these cases,
the data event and the parameter event are a priori statistically independent. For
example, the joint prior probability of $(\theta_{3},1)$ is $\pi(\theta
_{3})p(1\,|\,\theta_{3})=(1/3)(1/3)=\pi(\theta_{3})m(1)$ and similarly, the
joint prior probability of $(\theta_{1},2)$ is $\pi(\theta_{1})p(2\,|\,\theta
_{1})=(1/3)(1/3)=\pi(\theta_{1})m(2).$ For further discussion concerning this,
see Example 2.5. $\blacksquare$\smallskip

While there are natural reasons to consider the relative belief ratio as the
numerical measure of evidence, the principle of evidence implies that the
inferences made are largely independent of the choice of any valid measure of
evidence. For by the principle of evidence, there is evidence
in favor of $\psi$ iff $\pi_{\Psi}(\psi\,|\,x)>\pi_{\Psi}(\psi)$  and so all valid measures of evidence will agree that
there is evidence in favor in such a case. This implies that the plausible region $Pl_\Psi(x)$ is completely independent of the valid measure of evidence used. Notice that this can also be expressed, by (\ref{rb11}), as there is evidence in favor of 
$\psi$ iff $m(x\,|\,\psi)>m(x)$. This formulation is in terms of the likelihood function on the left and the average likelihood on the right, where the average is taken with respect to a given prior. If we accept the principle of evidence as an axiom in the pure probabilistic context, and this has broad acceptance, then this has an immediate implication for any approach to quoting an estimate and an interval for assessing the accuracy of that estimate, no matter how this is carried out. This can be stated as a theorem of inference. \medskip
\newline 
\noindent \textbf{Theorem 1.} Based upon inference base $I_{\Psi}=(m(\cdot\,|\,\psi):\psi\in\Psi(\Theta)\},x)$, the principle of evidence implies that the estimate with the most evidence in its favor is the MLE and any interval quoted, as part of assessing the accuracy of this estimate, must be a likelihood interval. 
\newline
\noindent Proof: This follows by adding any proper prior for $\psi$ to the inference base and then applying the principle of evidence to obtain the plausible region or a relative belief credible region within it.
$\blacksquare$
\medskip

There are several possible complaints concerning Theorem 1. One is that it is silent about which interval to quote. It is reasonable to answer, however, that this is not a problem when a prior on $\psi$ is provided, as with relative belief, and so this is a  problem that other approaches to inference have to deal with. A more serious concern is how to obtain the inference base  $I_{\Psi}$ for a marginal parameter? This is not a problem, however, when $\Psi(\theta)=\theta$, as with the Poisson with known background discussed in Section 3.1. This is also not a problem for relative belief because nuisance parameters are just integrated out.  Other approaches to this problem are conceivable but profiling is not one of these as, in general, it does not produce a valid likelihood. Some form of empirical Bayes, where a prior is selected from a family of priors based on the data, is conceivably a solution. For this, the data needs to be split into two parts with one part used to choose the hyperparameters of the family and the second part used for the inference step. This is, however, a topic for current research. Irrespective of these complaints it is clear that the FCCI does not satisfy Theorem 1 and this is also the case for many confidence regions and Bayesian credible regions. Note that it is reasonable to claim that any interval that does not satisfy Theorem 1 will generally include values for which there is evidence against them being the true value, see Section 3.1. 
\smallskip

\subsection{Checking the Ingredients}

The model and prior are both choices made by an investigator and so these are
subjective in nature. The ingredients of 
$I^{Bayes}=(\{p(\cdot\,|\,\theta):\theta\in\Theta\},\pi,x)$ need to be checked against the data to ensure that
reasonable choices have been made. Basically this means that the observed $x$ must be
reasonable for at least one of the distributions in the model. There are a
number of methods available for model checking and in general these can be
based on the conditional distribution of the data $x$, given a minimal
sufficient statistic (mss) $T(x),$ e.g., is the observed data in the tails of
this distribution, or is the observed value of an ancillary statistic $A(x)$
in the tails of its distribution?

Checking for prior-data conflict can also be carried out in a number of ways,
but here we employ the method presented in Evans and Moshonov (2006). The
approach is to compute the tail probability
\begin{equation}
M_{T}(m_{T}(t)\leq m_{T}(T(x))) \label{priorcheck}%
\end{equation}
where $T$ is a mss for the model $\{p(\cdot\,|\,\theta):\theta\in\Theta\}$
with prior measure $M_{T}$ and prior density $m_{T}(t)=\int_{\Theta}%
p_{T}(t\,|\,\theta)\pi(\theta)\,d\theta,$ where $p_{T}(\cdot\,|\,\theta)$ is
the density of $T$ when $\theta$ is true. It is proven in Evans and Jang
(2011a), under quite general conditions, that (\ref{priorcheck}) converges to
$\Pi(\pi(\theta)\leq\pi(\theta_{true}))$ with increasing amounts of data. So,
a small value of (\ref{priorcheck}) is indicating that the true value lies in
the tails of the prior and the prior may need to be changed. While
computing (\ref{priorcheck}) can be difficult, the following result provides
easier to compute bounds on (\ref{priorcheck}).\smallskip

\noindent\textbf{Proposition 1.} Denoting the MLE of $\theta$ by $\theta(t)$ when $t$ is observed,
then
\begin{equation}
M_{T}(p_{T}(t\,|\,\theta(t))\leq m_{T}(T(x)))\leq M_{T}(m_{T}(t)\leq
m_{T}(T(x)))\leq m_{T}(T(x)) \label{priorcheck2}%
\end{equation}
Proof: We have $m_{T}(t)=\int_{\Theta}p_{T}(t\,|\,\theta)\pi(\theta
)\,d\theta\leq p_{T}(t\,|\,\theta(t))$ which gives the LHS of
(\ref{priorcheck2}). For the RHS, by Markov's inequality,
\[
M_{T}(m_{T}(t)\leq m_{T}(T(x)))=M_{T}(1/m_{T}(t)\geq1/m_{T}(T(x)))\leq
m_{T}(T(x)).\text{ }\blacksquare
\]
Often $p_{T}(t\,|\,\theta(t))$ is available in closed form then, once
$m_{T}(T(x))$ has been computed, the LHS of (\ref{priorcheck2}) can be easily
computed via simulation and, if it is large, there is no prior-data conflict.
Also, if $m_{T}(T(x))$ is very small, it is immediate from (\ref{priorcheck2}%
), that there is prior-data conflict.

Evans and Jang (2011b) provides methodology for changing a prior in a
situation where there is prior-data conflict. Al
Labadi and Evans (2017) prove that relative belief inferences about $\psi$ are
optimally robust, among all Bayesian methodologies, to the prior $\pi_{\Psi}.$
It is also shown there, however, that when there is prior-data conflict, even
these inferences are not robust to the choice of $\pi_{\Psi},$ which
demonstrates the importance of this check. Nott et al. (2021) contains an
alternative approach to checking for prior-data conflict that is closely
connected to the relative belief ratio.

\subsection{Biases for Relative Belief and Frequentism}

A\ natural question to ask about an inference is, how reliable is it? This is the issue that frequentism addresses as one
considers how an inference will perform, with respect to error, in a conceptual series of repeats of
basically identical situations, with the data varying according to a
distribution in the model. If an inference procedure performs poorly, relative
to others, then one possibility is to replace it by a better performer or, if
the inference method is sound, increase the amount of data on which it is
based as generally this reduces error. This latter approach requires that the inference method be consistent,
namely, converge to the truth as the amount of data increases. Certainly,
relative belief inferences are consistent as can be seen from the close
association with likelihood methods. 

Requiring suitable performance under
repeated sampling is also part of the relative belief approach. The associated
calculations are a priori and are referred to as bias calculations which
measure errors under repeated sampling. The idea is that these calculations
are used as part of experimental design to ensure that sufficient data is collected
to produce reliable inferences, although they can also be used as part of a post
hoc analysis to assess the value of a study. A much more detailed discussion
of bias can be found in Evans and Guo (2021).

Consider first the \textbf{H} problem where there are two relevant bias
calculations. The \textit{bias against} $H_{0}:\Psi(\theta)=\psi_{0}$ is the a
priori probability,
\begin{equation}
\text{bias against}_{\Psi}(\psi_{0})=M(RB_{\Psi}(\psi_{0}\,|\,X)\leq
1\,|\,\psi_{0}), \label{hyp1}%
\end{equation}
where $M(\cdot\,|\,\psi_{0})$ denotes the conditional prior distribution of
the data given that $H_{0}$ is true, namely, the nuisance parameters have been
integrated out. If (\ref{hyp1}) is large, this indicates that not finding
evidence in favor of $H_{0},$ based on the observed data will happen with high
prior probability, when $H_{0}$ is true. As such, it cannot be claimed that
finding evidence against is a reliable inference in such a context. The bias
in favor of $H_{0}$ requires a metric $d$\ on $\Psi(\Theta)$ together with a
deviation $\delta$ such that, if $d(\psi,\psi_{0})<\delta,$ then we don't
regard the difference between $\psi$ and $\psi_{0}$ to be of scientific
importance. So, $d$ and $\delta$ are determined by the application. Then the
\textit{bias in favor of} $H_{0}$ is the prior probability%
\begin{equation}
\text{bias in favor}_{\Psi}(\psi_{0})=\sup_{\psi:d(\psi_{0},\psi)\geq\delta
}M(RB_{\Psi}(\psi_{0}\,|\,X)\geq1\,|\,\psi). \label{hyp2}%
\end{equation}
If (\ref{hyp2}) is large, then there is a large prior probability of not
obtaining evidence against $H_{0}$ when it is meaningfully false. As such,
finding evidence in favor of $H_{0}$ in such a context, is not a reliable
inference. Typically $M(RB_{\Psi}(\psi_{0}\,|\,x)\geq1\,|\,\psi)$ decreases as
$\psi$ moves away from $\psi_{0},$ so it is only necessary to consider values
of $\psi$ satisfying $d(\psi_{0},\psi)=\delta$ in (\ref{hyp2}) to determine
the bias in favor. Notice that (\ref{hyp1}) and (\ref{hyp2}) can be considered
as frequentist probabilities for the model given by $\{m(\cdot\,|\,\psi
):\psi\in\Psi\}$ which corresponds to the full model when $\Psi(\theta
)=\theta.$ Clearly, these are relevant error probabilities.

For the \textbf{E} problem we consider the average bias against a value of
$\psi\sim\pi_{\Psi},$ so%
\begin{align}
\text{bias against}_{\Psi}  &  =E_{\Pi_{\Psi}}(M(RB_{\Psi}(\psi\,|\,X)\leq
1\,|\,\psi))=E_{\Pi_{\Psi}}(M(\psi\notin Pl_{\Psi}(X)\,|\,\psi))\nonumber\\
&  =1-E_{\Pi_{\Psi}}(M(\psi\in Pl_{\Psi}(X)\,|\,\psi)), \label{cov1}%
\end{align}
which is the prior probability that the plausible region doesn't contain the
true value. Note that $E_{\Pi_{\Psi}}(M(\psi\in Pl_{\Psi}(x)\,|\,\psi))$ is
the prior probability that the plausible region contains the true value and
can be considered as the average confidence level of $Pl_{\Psi}(x)$ with
respect to the model $\{m(\cdot\,|\,\psi):\psi\in\Psi\},$ where the average is
with respect to the prior on $\psi.$ Typically an upper bound can be obtained
for $M(RB_{\Psi}(\psi\,|\,x)\leq1\,|\,\,\psi)$ as a function of $\,\psi.$ In
that case, 1 minus this bound serves as the frequentist confidence of
$Pl_{\Psi}(x)$ with the respect to the model $\{m(\cdot\,|\,\psi):\psi\in
\Psi\}.$

The average bias in favor is%
\begin{align}
\text{bias in favor}_{\Psi}  &  =E_{\Pi_{\Psi}}\left(  \sup_{\psi_{0}%
:d(\psi,\psi_{0})\geq\delta}M(RB_{\Psi}(\psi_{0}\,|\,X)\geq1\,|\,\psi)\right)
\nonumber\\
&  =E_{\Pi_{\Psi}}\left(  \sup_{\psi_{0}:d(\psi,\psi_{0})\geq\delta}M(\psi
_{0}\notin\mbox{$Im_\Psi(X)$}\,|\,\psi)\right)  . \label{cov2}%
\end{align}
This is the prior probability that a meaningfully false value is not in the
\textit{implausible region} $\mbox{$Im_\Psi(x)$}=\{\psi:RB_{\Psi}%
(\psi\,|\,x)<1\},$ the set of values for which there is evidence against.

For the \textbf{H} and \textbf{E} problems it is desirable to have both biases
small as these quantities refer to errors in the inference.\ As discussed in
Evans (2015), both biases converge to 0 with increasing sample size. As such,
the biases are used in the design phase of a study to ensure that the results
obtained will be reliable.

Notice that the biases only depend on the principle of evidence. So, if
any other valid measure of evidence is used, the bias calculations are the
same. Bias against and bias in favor are playing similar roles to size and
power in frequentist hypothesis testing problems and to confidence and false
coverage in frequentist estimation problems. It is also to be noted that,
choosing a prior to reduce one form of bias, typically raises the other. The
aim should be to have both biases at reasonable values and this is controlled
by sample size. As with Proposition 2, it is possible to compute bounds on
the biases to help to cut down on computation.

The following example makes an important point about when no evidence is
obtained either way. It is placed in this section because it has implications for error control via coverage probabilities. \smallskip

\noindent\textbf{Example 2.5.}

Suppose now that the model is as given in Table 4 with $\mathcal{X}=\{1,2,3\}$
and suppose the prior is the uniform prior on $\Theta=\{\theta_{1},\theta
_{2},\theta_{3}\}.$%

%TCIMACRO{\TeXButton{B}{\begin{table}[H] \centering}}%
%BeginExpansion
\begin{table}[H] \centering
%EndExpansion%
\begin{tabular}
[c]{c|ccc}
& $p(1\,|\,\theta)$ & $p(2\,|\,\theta)$ & $p(3\,|\,\theta)$\\\hline
$\theta_{1}$ & \multicolumn{1}{|l}{$1/2$} & \multicolumn{1}{l}{$499/1000$} &
\multicolumn{1}{l}{$1/1000$}\\
$\theta_{2}$ & \multicolumn{1}{|l}{$1/1000$} & \multicolumn{1}{l}{$998/1000$}
& \multicolumn{1}{l}{$1/1000$}%
\end{tabular}
\caption{Model for Example 2.5.}\label{TableKey copy(1)}%
%TCIMACRO{\TeXButton{E}{\end{table}}}%
%BeginExpansion
\end{table}%
%EndExpansion

\noindent Therefore, $Pl(1)=\{\theta_{1}\},Pl(2)=\{\theta_{2}\},$ as there is
evidence in favor in both cases, and $Pl(3)=\phi$ as no evidence in favor of
either value is found when $x=3.$ Note that with this assignment for $Pl(3),$
there is no contribution to the coverage or false coverage probabilities, namely, the biases, when
$x=3$, as indeed there shouldn't be. Observing $x=3$ tells us nothing about the true
value. This shows that it is possible for the data to lead to no evidence
either way but this only happens when the likelihood is constant in $\theta.$
More important is noting that the proper expression of this is via the null
set, as some have argued that recording the entire parameter space is the
appropriate approach to saying no evidence has been obtained. But that is not
correct, as it is impossible for every value of $\theta$ to have evidence in
its favor (or against) when the model has been accepted as being correct. As has already been mentioned, the plausible region can
never be equal to the whole parameter space. $\blacksquare$ \smallskip

The overall message of this section is that frequentist error concerns are highly relevant for relative belief. Moreover, the plausible region avoids many of the problems associated with the construction of confidence regions as discussed in Evans et al. (2024). Overall, there is a net gain from the point-of-view of frequentism, in the use of relative belief inferences.

\section{Application to Poisson-based Inference}

Suppose interest is in making inference about the rate $\lambda$ at which new
particles are being produced in an experiment where there is a background rate
$b$ of noise particles, so the overall rate of particle production is
$\theta=\lambda+b.$ The data is $x=(x_{1},\ldots,x_{n}),$ where $x_{i}$ is the
number of events observed in the $i$-th experiment, and the time for
observing the number of events is the same in each experiment. It is supposed
that the data arises as a sample of independent values from a Poisson$(\lambda
+b)$ distribution where $\lambda,b\in\lbrack0,\infty).$ For this model a
minimal sufficient statistic is given by the total $T(x)=\sum_{i=1}^{n}%
x_{i}=n\bar{x}\sim$ Poisson$(n(\lambda+b))$ and we can disregard all aspects
of the data beyond $T(x)$ for the purposes of inference. The basic model is
then $\{p_{T}(\cdot\,|\,\lambda,b):\lambda,b\in\lbrack0,\infty)\}$ where
$p_{T}(t\,|\,\lambda,b)=n^{t}(\lambda+b)t^{t}e^{-n(\lambda+b)}/t!$ for $t\in%
%TCIMACRO{\U{2115} }%
%BeginExpansion
\mathbb{N}
%EndExpansion
_{0}.$

In this situation interest is in making inference about $\lambda$ and two
scenarios are considered, namely, $b$ is known, perhaps based on previous
experiments, and $b$ is unknown. In the latter context it is clear that the
model suffers from nonidentifiability as $\lambda$\ and $b$ cannot be
separated based on the data alone. In the first situation a prior is placed on
$\lambda.$ In the second context, independent priors are placed on both
$\lambda$ and $b$ and this allows for the nuisance parameter $b$ to be
integrated out leaving a marginal model for $\lambda.$

The data beyond the value $T(x)$ is useful for model checking and, in
particular,%
\[
(x_{1},\ldots,x_{n})\,|\,T(x)\sim\text{ multinomial}(T(x),1/n,\ldots,1/n).
\]
So, if the counts $x_{i}$ are surprising relative to this distribution, then
this is an indication that the Poisson model is wrong. For example, it could
be that the rate $\lambda+b$ is changing with the experiment conducted and
there could be other kinds of failure.

\subsection{Background Rate is Known}

To properly express the evidence in the data about $\lambda$, it is necessary
to specify a prior. A convenient choice is a gamma$_{\mathrm{{rate}}}(\tau
_{1},\tau_{2})$ distribution as the two degrees of freedom associated with the
hyperparameters $(\tau_{1},\tau_{2}),$ allows for a wide expression of
beliefs. The hyperparameters are elicited based on what is understood about
the experimental outcomes a priori. Since $\lambda$ represents the mean number
of new particle events, an interval $(l_{1},u_{1})$ is specified such that it
is believed that $\lambda\in(l_{1},u_{1})$ with virtual certainty, namely,
with a probability $\gamma$ close to 1, such as $\gamma=0.99.$ Note that we
could take $l_{1}=0$ but this is not necessary. In addition, the mode
$m=(\tau_{1}-1)/\tau_{2}\in(l_{1},u_{1})$ is specified so $\tau_{1}%
=1+m\tau_{2}$ and then $\tau_{2}$ is found such that $\gamma=G(u_{1};\tau
_{1},\tau_{2})-G(l_{1};\tau_{1},\tau_{2}),$ where $G(\cdot;\tau_{1},\tau_{2})$
is the cdf of the gamma$_{\mathrm{{rate}}}(\tau_{1},\tau_{2})$ distribution
with density $g(\cdot;\tau_{1},\tau_{2})$. For example, if the values
$(l_{1},u_{1})=(0,10),m=1.5$ and $\gamma=0.99$ are specified, then the
gamma$_{\mathrm{{rate}}}(1.975,0.65)$ prior is determined. Rather than a
general gamma prior for $\lambda$, perhaps a more suitable prior is a
gamma$_{\mathrm{{rate}}}(1,\tau_{2})$ as this has its mode at 0 and is a
conservative choice as more weight is placed on the nonexistence of a new
particle being found.. In this case, it makes sense to take $l_{1}=0$ and then
$\tau_{2}=\log(1-\gamma)/l_{2}.$

It is necessary to check the prior based on the observed data, to make sure
that it places sufficient mass near the true value. To implement the check for
prior-data conflict (\ref{priorcheck}), requires computing, for each $t$,%
\begin{align*}
m_{T}(t)  &  =\int_{0}^{\infty}\frac{n^{t}(b+\lambda)^{t}}{t!}\exp\left\{
-n(b+\lambda)\right\}  g(\lambda;\tau_{1},\tau_{2})\,d\lambda\\
&  =\left\{  \frac{(nb)^{t}}{t!}e^{-nb}\right\}  \left(  \frac{\tau_{2}%
}{n+\tau_{2}}\right)  ^{\tau_{1}}\int_{0}^{\infty}\left(  1+\frac{u}%
{(n+\tau_{2})b}\right)  ^{t}g(u;\tau_{1},1)\,du,
\end{align*}
which is a Poisson$(nb)$ probability function times a nonnegative function of
$b.$ The check, (\ref{priorcheck}) then equals
$1-\sum_{\{t:m_{T}(t)>m_{T}(T(x))}m_{T}(t).$ When $\tau_{1}=1,$ then
$m_{T}(t)$ simplifies to
\[
m_{T}(t)=\left(  \frac{n}{n+\tau_{2}}\right)  ^{t}\left(  \frac{\tau_{2}%
}{n+\tau_{2}}\right)  e^{\tau_{2}b}(1-G((n+\tau_{2})b;t+1,1),
\]
where the product of the first two terms specifies a geometric$(\tau
_{2}/(n+\tau_{2})$ distribution on $%
%TCIMACRO{\U{2115} }%
%BeginExpansion
\mathbb{N}
%EndExpansion
_{0}.$ If (\ref{priorcheck}) is small, it suggests that the true value of
$\theta=b+\lambda$ lies outside the effective support for $\theta$ induced by
the prior. This can be avoided by decreasing $\tau_{2}$ as this induces more
spread in the prior, see Evans and Jang (2011b). When $b=0,$ then
\begin{equation}
m_{T}(t)=\left(  \frac{n}{n+\tau_{2}}\right)  ^{t}\left(  \frac{\tau_{2}%
}{n+\tau_{2}}\right)  \frac{\Gamma(t+\tau_{1})}{\Gamma(t+1)\Gamma(\tau_{1})}
\label{rbcheck2}
\end{equation}
and, when $\tau_{1}=1,$ this is the geometric$(\tau_{2}/(n+\tau_{2})$ distribution.

The relative belief ratio for $\lambda$ equals
\[
RB(\lambda\,|\,x)=\frac{n^{T(x)}(b+\lambda)^{T(x)}e^{-n(b+\lambda)}%
/T(x)!}{m_{T}(T(x))}%
\]
which implies that the MLE of $\lambda$ is $\lambda(x)=T(x)/n-b$ when $T(x)/n\ge b$, and $\lambda(x)=0$ otherwise.  This
converges wp1 to $\lambda_{true}$ with increasing sample size$.$ The
plausible interval for $\lambda$ is given by%
\[
Pl(x)=\{\lambda:RB(\lambda\,|\,x)>1\}=(\lambda_{1}(x),\lambda_{2}(x))
\]
where the $\lambda_{i}(x)$ satisify $RB(\lambda_{i}(x)\,|\,x)=1$ and, based on
the asymptotics of the likelihood, this shrinks to $\{\lambda_{true}\}$ with
increasing sample size.

The biases for $H_{0}:\lambda=\lambda_{0}$ are given by%
\begin{align*}
\text{bias against}(\lambda_{0})  & =M_{T}(RB(\lambda_{0}\,|\,t)\leq
1\,|\,\lambda_{0})=1-\sum_{\{t:RB(\lambda_{0}\,|\,t)>1\}}\frac{n^{t}%
(b+\lambda_{0})^{t}}{t!}e^{-n(b+\lambda_{0})},\\
\text{bias in favor}(\lambda_{0})  
& =\sup_{\lambda:|\lambda_{0}-\lambda
|\geq\delta/2}M_{T}(RB_{\Psi}(\lambda_{0}\,|\,t)\geq1\,|\,\lambda) \\
& =\sup
_{\lambda\in\{\lambda_{0}-\delta/2,\lambda_{0}+\delta/2\}}\sum_{\{t:RB(\lambda
_{0}\,|\,t)\geq1\}}\frac{n^{t}(b+\lambda)^{t}}{t!}e^{-n(b+\lambda)},
\end{align*}
where $\delta$ is the difference that matters. For the estimation problems the
biases are obtained by averaging these biases over $\lambda_{0}$ with respect
to the prior (now placed on $\lambda_{0}).$

Consider now implementing this in a numerical example. \smallskip

\noindent\textbf{Example 3.1.} \textit{Background $b$ is known.}

To illustrate the use of the relative belief approach, consider a simulated example where $b=1$ is smaller than $\lambda=5$, so the observed count is generated from a Poisson$(6)$. It is supposed that our inferential goal is to assess the hypothesis $H_0 : \lambda=0$ as well as provide the estimate and plausible interval for the true value of $\lambda$. 

The prior for $\lambda$ is elicited by providing an interval that it is believed contains the true $\lambda$ with reasonably high prior probability. In this case the interval $(0,10)$, having prior probability content $0.95$, was specified. Using a gamma$(1,\tau)$ prior, as this has its maximum at 0, leads to the choice $ \tau= 0.30$.
 
The next step is to determine an appropriate sample size $n$ to control the biases a priori. Note that this is effectively controlling the error rates in repeated sampling. Table 5 contains the lower bound for the frequentist confidence for the plausible interval, the Bayesian confidence for the plausible interval, and the bias against the hypothesis $H_0$ for various sample sizes $n$. Table 6 contains the bias in favor calculations for various sample sizes and meaningful differences $\delta$. 

From Table 5 it is seen that $n=10$ measurements are needed to ensure $90\%$ frequentist confidence for the plausible interval while the Bayesian confidence is $0.943$.
So, good coverage is attained for relatively small sample sizes. 
The bias against $H_0$ is $0.049$, which attains the conventional $0.05$ type I error bound. The bias in favor calculations in Table 6 reveal, as is typical, that with decreasing values of the meaningful difference $\delta$, much higher sample sizes are required to make the bias in favor small. Recall that for $H_0$, the bias in favor is the prior probability of getting evidence in favor when it is meaningfully false, so 1 minus this quantity is just like power in the frequentist setting.
 
\begin{table}[t]
\centering

\begin{tabular}{lccc}

$n$ & Frequentist Conf. & Bayesian Conf. & $H_0 :\lambda=0$  \\
\hline
1 &  0.703 & 0.839 & 0.080   \\
5 &  0.860 & 0.920 & 0.068   \\
10 &  0.904  & 0.943 & 0.049   \\
25 &  0.943  & 0.965 & 0.023   \\
50 &  0.960  & 0.976 & 0.017   \\
100 & 0.974  & 0.983 & 0.011   \\
\hline
\end{tabular}
\caption{Frequentist confidence and Bayesian confidence for the plausible interval, and bias against for $H_0$, for several values of $n$ in Example 3.1.}
\label{tab:bias_against}
\end{table}

\begin{table}[h!]
\centering
\begin{tabular}{lcc}
\multicolumn{3}{c}{$\delta = 1$} \\
\hline
$n$ & $H_0$ & Bif \\
\hline
1 & 0.677 & 0.823 \\
5 & 0.333 & 0.695 \\
10 & 0.156 & 0.576 \\
25  & 0.016 & 0.369 \\
50   & 0.000 & 0.204 \\
100  & 0.000 & 0.085 \\
\hline
\end{tabular}
\hspace {.5cm}
\begin{tabular}{lcc}

\multicolumn{3}{c}{$\delta = 2$} \\
\hline
$n$ & $H_0$ & Bif \\
\hline
1 & 0.423 & 0.625 \\
5 & 0.037 & 0.380 \\
10 & 0.002 & 0.221 \\

25  & 0.000 & 0.076 \\
50  & 0.000 & 0.021 \\
100  & 0.000 & 0.003 \\

\hline
\end{tabular}
\hspace{.5cm}
\begin{tabular}{lcc}

\multicolumn{3}{c}{$\delta = 3$} \\
\hline
$n$ & $H_0$ & Bif \\
\hline
1 & 0.238 & 0.509 \\
5 & 0.002 & 0.192 \\
10 & 0.000 & 0.084 \\

25  & 0.000 & 0.016 \\
50  & 0.000 & 0.002 \\
100  & 0.000 & 0.000 \\

\hline
\end{tabular}
\caption{Bias in favor for $H_0$ and the plausible interval (Bif) for several values of $n$ and  $\delta$ in Example 3.1.} 
\label{tab:bias_in_favor_split}
\end{table}

 A generated sample of size $n = 10$, from the Poisson$(6)$ distribution produced a total number of $T(x) = 57$ events, so $\bar{x} = 5.7$ and $\lambda(x)=4.70$ is the MLE. The prior predictive check (4) equals $0.243$ and so there is no significant conflict between the prior and the data.
 
 Figure 1 is a plot of the prior, posterior and relative belief ratio for $\lambda$. For assessing $H_0 : \lambda=0$, then $RB(0 \, | \, x)=0.00$. By Proposition 9 in Al-Labadi et al. (2025),  $RB(0 \, | \, x)$ serves as an upper bound on   the strength and so this is also equal to 0.00. As such, this is extremely strong evidence against $H_0$. The plausible interval for $\lambda$ is $Pl(x)=(3.33, 6.34)$ and this has posterior probability content $0.954$. So, there is a high degree of belief that the true value is in the plausible interval (which it is, as $\lambda_{true}=5$), but the length indicates a reasonable degree of uncertainty that the MLE $\lambda(x)=4.70$ is close to the true value. As the bias in favor calculation illustrates, the prior probability of the plausible interval containing a meaningfully false value is not small at $0.221$ when $n=10, \delta=2$, so this is not surprising.

\begin{figure}[t]
    \centering
    \includegraphics[width=1\linewidth]{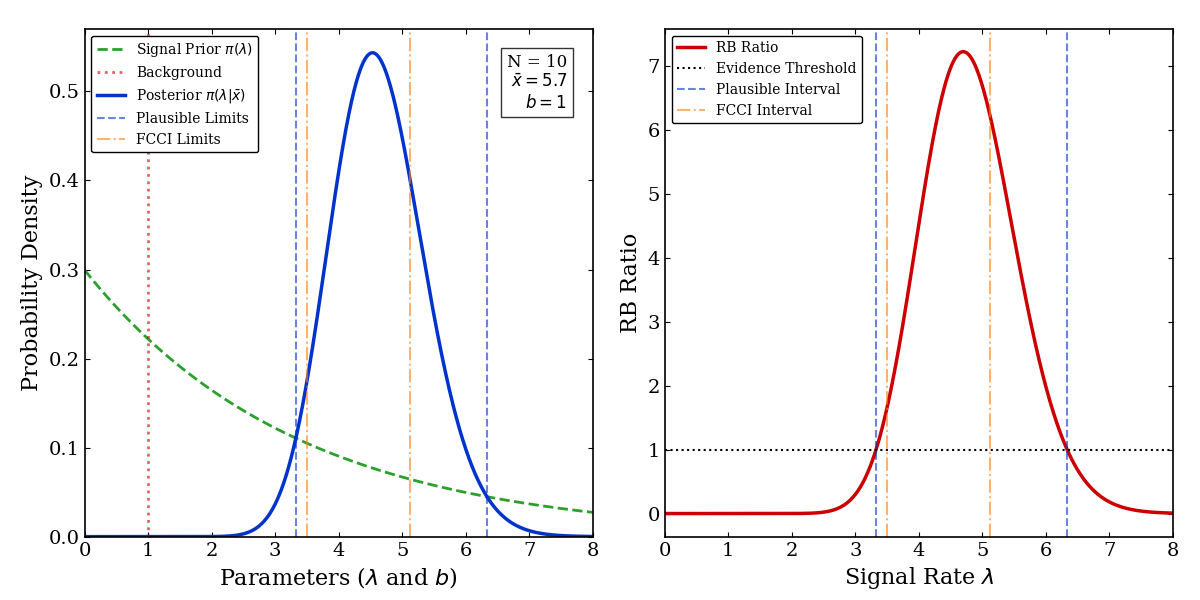}
    \caption{Plots for Example 3.1 based on a generated sample of $n=10$. Left panel: Prior density, background, posterior density, limits for FCCI and plausible interval for $\lambda$. Right panel: Relative belief ratio for $\lambda$ where the horizontal dashed line at 1 marks the evidence cutoff, the FCCI and plausible interval.}
    \label{fig:example31}
\end{figure} 

 The FCCI matching the frequentist coverage (0.904) of $Pl(x)$ is $(3.50, 5.13)$. While this is much shorter than the plausible interval, it excludes values for which there is evidence in favor. Also $RB(3.50 \,|\,x)=1.66 \neq RB(5.13 \,|\,x)=6.24 $, so the FCCI violates the likelihood ordering quite strongly.
 Recall that this contradicts what the data is indicating, namely, there is more support from the data, whether via evidence or likelihood,  for values that are outside FCCI than for some of the values inside. $Pl(x)$ provides a posterior belief of $0.954$ that the true value is inside, while the posterior content of the FCCI is $0.699$. The $0.699$-relative belief credible region is $(3.96, 5.51)$, with $RB(3.96 \,|\,x)= RB(5.51 \,|\,x)=4.25$ so, as always, it is a likelihood interval, and it is a bit shorter than the FCCI with this posterior content. At least in this example, the plausible interval, based on a fairly weakly informative prior, gives a better summary of the evidence in the data concerning the true value.  $\blacksquare$

\smallskip
In the following example a more comprehensive comparison is made between the FCCI and the plausible interval.
\smallskip

\noindent\textbf{Example 3.2.} \textit{Simulation for comparison with FCCI.}

For this example the same model and prior is used as in Example 3.1. Now, however, we consider how the FCCI, specified by the frequentist confidences for the plausible interval in Table 5,  performs when the data is repeatedly generated from the prior predictive. 

Table 7 records the prior probability that the FCCI will contain values that have evidence against them according to the principle of evidence, and the prior probability that the FCCI violates the likelihood ordering. These probabilities are relatively high and it is notable that the plausible interval does not suffer from such violations. 

Table 8 records the prior probabilities that the FCCI contains the meaningfully false values $\lambda_{true} \pm \delta/2$ for several values of $n$ and $\delta$. The entries in this table are directly comparable with the Bias in favor (Bif) column of Table 5 and we see that the plausible interval performs much better on this criterion. Recall that bias against is associated with errors of the first kind (finding evidence against when something is true), which in the case of a reported interval means the interval doesn't contain the true value, and bias in favor is associated with errors of the second kind (finding evidence in favor when something is false), which in the case of a reported interval means the interval contains meaningfully false values. $\blacksquare$ 

% --- Table 1: Coverage and LR diagnostics ---
\begin{table}[t]
\centering
\label{tab:coverage_rb}
\begin{tabular}{lcccc}
\toprule
$n$ & Confidence & Prob. violates evidence & Prob. violates likelihood ordering \\
\midrule
1 & 0.703 & 0.485 & 1.000 \\
5 & 0.860  & 0.407 & 1.000 \\
10 & 0.904 & 0.290 & 1.000 \\
25 & 0.943  & 0.257 & 0.997 \\
50 & 0.960 & 0.076 & 0.999 \\
100 & 0.974 & 0.071 & 0.974 \\
\bottomrule
\end{tabular}
\caption{Prior probabilities that FCCI violates evidence (includes values that have evidence agianst them) and violates the likelihood ordering (excludes values which have higher likelihoods than values included) in Example 3.2. }
\end{table}

% --- Table for δ = 1.0 ---
\begin{table}[t]
\centering
\label{tab:type_ii_delta_1}
\begin{tabular}{lccc}
\toprule
$n$ & Coverage $\delta=1$ & Coverage $\delta=2$ & Coverage $\delta=3$\\
\midrule
1 & 0.827 & 0.844 & 0.799 \\
5 & 0.934 & 0.848 & 0.647 \\
10 & 0.943 & 0.756 & 0.439 \\
25 & 0.916 & 0.472 & 0.158 \\
50 & 0.803 & 0.226 & 0.041 \\
100 & 0.575 & 0.071 & 0.005 \\
\bottomrule
\end{tabular}
\caption{Coverage of at least one of the meaningfully false values $\lambda_{true} \pm \delta/2$ for several values of $n$ and $\delta$ in Example 3.2.}
\end{table}

\smallskip
The plausible interval and the FCCI are compared with respect to a commonly noted strange behavior of the FCCI in the following example. 
\smallskip
\newline
\noindent\textbf{Example 3.3.} \textit{Sensitivity to choice of $b$ when 0 events are observed.}

One odd feature of a FCCI arises when no counts are observed, as is surely possible. It has been noted that, in such a context, the length of the FCCI for $\lambda$ decreases as $b$ increases. This doesn't really make sense because $b$ and $\lambda$ are independent. The upper limit should remain unchanged regardless of $b$ in such a context. Figure 2 demonstrates this for two sample sizes with the confidences as specified in Table 5 and the weakly informative prior specified in Example 3.1. Both the FCCI and the plausible interval have the left-hand end-point equal to 0. The plausible interval of $\lambda$ does not depend on $b$ for a given sample size. This is not the case for the FCCI although the effect is largely removed for the larger sample size. While the FCCI upper limit appears to be more accurate when $n=10$, it is not reporting values of $\lambda$  for which there is evidence in favor, so one could argue that it does not represent an appropriate assessment of accuracy here. $\blacksquare$

\begin{figure}[t]
    \centering
    \includegraphics[width=1\linewidth]{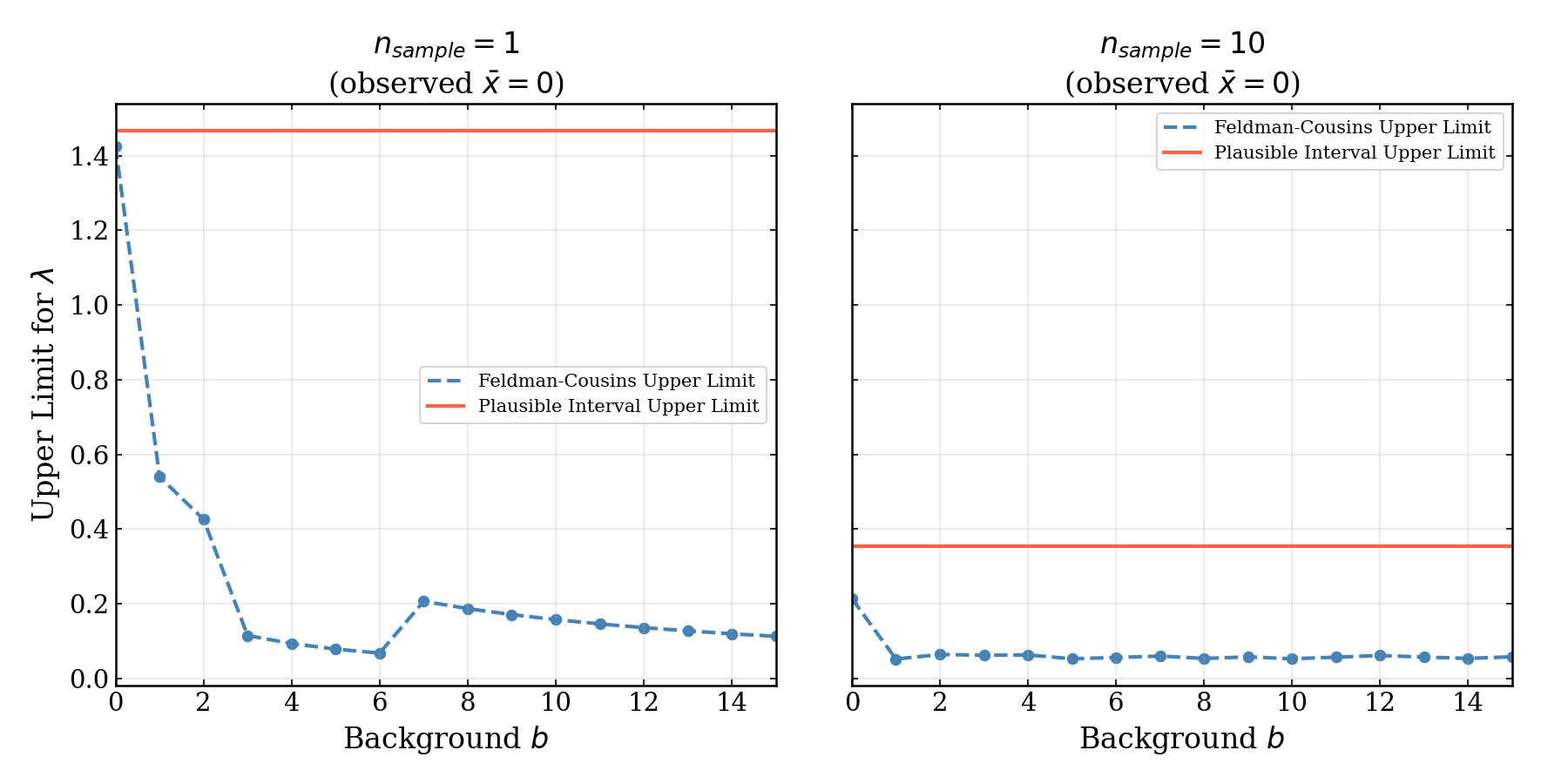}
    \caption{Plots for Example 3.3. With 0 counts observed, the FCCI upper limit fluctuates and generally decreases as the background $b$ increases, while the upper limit of the plausible  interval is invariant with respect to $b$.}
    \label{fig:example32}
\end{figure} 

\subsection{Background rate is unknown.}

The natural approach, when $b$ is unknown, to also place an independent prior on this quantity and treat it as a nuisance parameter. It is supposed
then that the prior structure is given by
\[
\lambda\sim\text{gamma}_{\mathrm{{rate}}}(\tau_{1},\tau_{2})\text{ independent
of }b\sim\text{gamma}_{\mathrm{{rate}}}(\varsigma_{1},\varsigma_{2})
\]
and the hyperparameters $(\tau_{1},\tau_{2})$ and $(\varsigma_{1},\varsigma_{2})$ are elicited as described in Example 3.1.
Inference about $\lambda$ is then based on the prior $\pi_{\Lambda}$ for
$\lambda$ and the model $\{m(\cdot\,|\,\lambda):\lambda\geq0\}$ where
\[
m_{T}(t\,|\,\lambda)=\int_{0}^{\infty}(n^{t}(b+\lambda)^{t}e^{-n(b+\lambda
)}/t!)\,\text{gamma}_{\mathrm{{rate}}}(b;\varsigma_{1},\varsigma_{2})\,db.
\]
The corresponding relative belief ratio is then
\[
RB_{\Lambda}(\lambda\,|\,x)=\frac{m_{T}(T(x)\,|\,\lambda)}{m_{T}(T(x))}%
\]
and inference proceeds as in the case where $b$ is known. The biases are now
computed using the ingredients $\{m(\cdot\,|\,\lambda):\lambda\geq0\}$ and
$\pi_{\Lambda}.$  So, conceptually there is little difference between the two contexts for inference about $\lambda$, although the computational burden is higher when $b$ is unknown due to the additional integration.

One place where the analyses differ is in the check for prior-data conflict. This is a check on the prior on $\theta=b + \lambda$ induced by the individual priors on $b$ and on $\lambda$. If the prior fails its check, however, it isn't clear if this failure is due to the prior on $b$, the prior on $\lambda$ or both. This may be the state of affairs in a given application, but one possible approach here is that practitioners first conduct an experiment dedicated to determine a suitable prior for $b$ when this is possible. The check on the prior on $b$ can then be implemented via (\ref{rbcheck2}) with $(\varsigma_1, \varsigma_2)$ replacing $(\tau_1,\tau_2)$ in that formula. Supposing, then that the prior on $b$ does not produce a conflict, a reasonable check on the prior on $\lambda$ is to set $b=(\varsigma_1-1)/\varsigma_2$, the mode of the gamma$(\varsigma_1,\varsigma_2)$ prior, and perform the check on this conditional prior on $\lambda$, as in Example 3.1. It is worth mentioning here too, that if the prior on $b$ is too dispersed relative to that on $\lambda$, then inference on $\lambda$ will be much less informative.  

A numerical example is now considered. 
\smallskip
\newline
\noindent\textbf{Example 3.4.} \textit{Background $b$ is unknown but relatively small.}

Consider first the situation of Example 3.1 where the same prior on $\lambda$ is used and with the same generated data. Recall that
$57$ events were observed based on a sample of size of $n=10$ from a Poisson$(6)$ with $b=1$. The prior on $b$ 
is taken to be elicited based on the interval $(0.5,1.5)$ containing $95\%$ of the prior probability. This assumes that $b$ is known relatively accurately but not precisely. With the mode at $1$ and the prior content of the interval as specified, this leads to a $\text{gamma}_{rate}(18.47,17.47)$ prior for $b$.

Accounting for the background nuisance parameter and assuming elicited priors are true, the bias against $H_0:\lambda=0$ is 0.057, and the bias in favor of $H_0$ (meaningful difference $\delta=2$) is 0.002 with a Bayesian statistical power of 0.998.
Based on the fact that the value $RB(0 \,|\,x) = 0.000$ provides an upper bound on the strength measurement, it can be concluded that categorical evidence against $H_0$ has been obtained. The resulting plausible interval for $\lambda$ is $(3.22, 6.32)$, and it contains $94.7\%$ of the posterior probability. It is notable that the inferences differ very little from the case where $b$ is known.
$\blacksquare$ \smallskip

\begin{table}[t]
\centering
\caption{Bias Against and Coverage}
\begin{tabular}{lccc}
\hline
n & Frequentist Conf. & Bayesian Conf. & $H_0 :\lambda=0$  \\
\hline
1 & 0.705 & 0.834 & 0.096   \\
5 & 0.856 & 0.915 & 0.066   \\
10 & 0.894 & 0.937 & 0.057   \\
25 & 0.928 & 0.955 & 0.042   \\
50 & 0.940 & 0.964 & 0.043   \\
100 & 0.948 & 0.969 & 0.037   \\
\hline
\end{tabular}
\label{tab:bias_against_vb_small_b}
\end{table}

\begin{table}[t]
\centering
\begin{tabular}{lcc}
\multicolumn{3}{c}{$\delta = 1$} \\
\hline
$n$ & $H_0$ & Bif \\
\hline
1 & 0.677 & 0.823 \\
5 & 0.333 & 0.695 \\
10 & 0.156 & 0.576 \\
25 & 0.016 & 0.369 \\
50 & 0.000 & 0.204 \\
100 & 0.000 & 0.085 \\
\hline
\end{tabular}
\hspace {.5cm}
\begin{tabular}{lcc}

\multicolumn{3}{c}{$\delta = 2$} \\
\hline
$n$ & $H_0$ & Bif \\
\hline
1 & 0.423 & 0.625 \\
5 & 0.037 & 0.380 \\
10 & 0.002 & 0.221 \\
25 & 0.000 & 0.076 \\
50 & 0.000 & 0.021 \\
100 & 0.000 & 0.003 \\
\hline
\end{tabular}
\hspace{.5cm}
\begin{tabular}{lcc}

\multicolumn{3}{c}{$\delta = 3$} \\
\hline
$n$ & $H_0$ & Bif \\
\hline
1 & 0.238 & 0.509 \\
5 & 0.002 & 0.192 \\
10 & 0.000 & 0.084 \\
25 & 0.000 & 0.016 \\
50 & 0.000 & 0.002 \\
100 & 0.000 & 0.000 \\
\hline
\end{tabular}
\caption{Bias in favor for $H_0$ and the plausible interval (Bif) for several values of $n$ and  $\delta$ in Example 3.4.} 
\label{tab:bias_in_favor_split_vb_small_b}
\end{table}

\begin{figure}[t]
    \centering
    \includegraphics[width=1\linewidth]{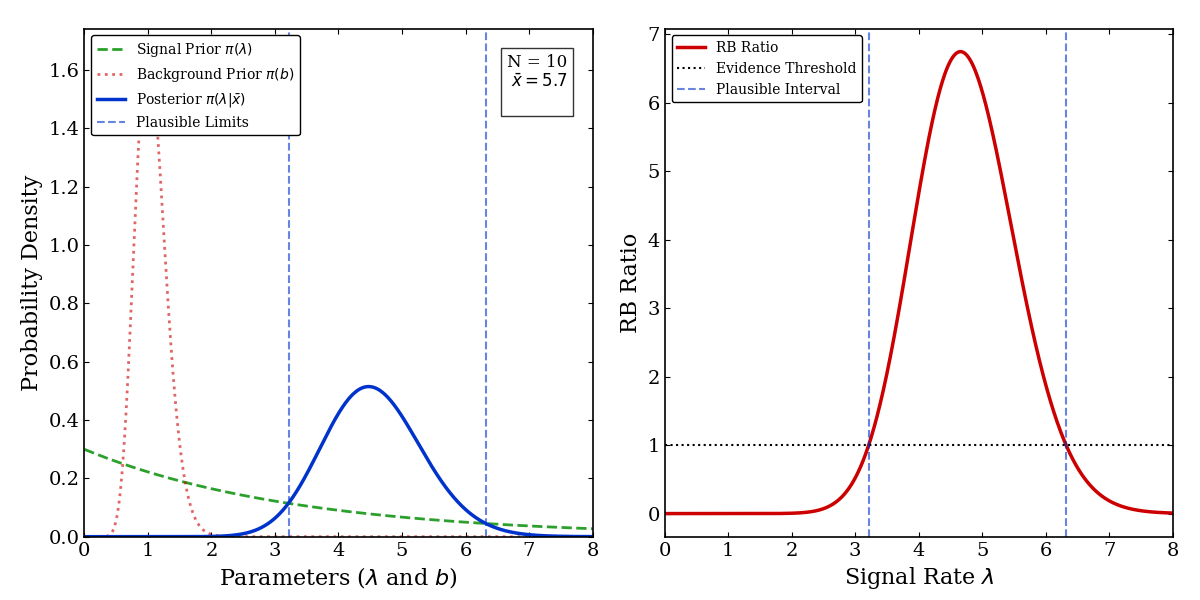}
    \caption{Plots for Example 3.4 based on a generated sample of $n=10$. Left panel: prior on $\lambda$, prior on $b$, posterior density, and plausible interval for $\lambda$. Right panel: Relative belief ratio for $\lambda$ where the horizontal dashed line at 1 marks the evidence cutoff and plausible interval.}
    \label{fig:example34}
\end{figure}

\noindent\textbf{Example 3.5.} \textit{Background $b$ is unknown and dominates.}

Consider now a background-dominated case where $b=5$ and $\lambda=1$. The same prior on $\lambda$ is used and with the same generated data, that is,
$57$ events were observed based on a sample of size of $n=10$ from a Poisson$(6)$ with $b=5$. The prior on $b$ 
is taken to be elicited based on the interval $(4.5,5.5)$ containing $95\%$ of the prior probability. This assumes that $b$ is known relatively accurately but not precisely. With the mode at $5$ and the prior content of the interval as specified, this leads to a $\text{gamma}_{rate}(387.17,77.23)$ prior for $b$.

Accounting for the background nuisance parameter and assuming elicited priors are true, the bias against $H_0:\lambda=0$ is $0.088$, and the bias in favor of $H_0$ (meaningful difference $\delta=2$) is $0.127$ with a Bayesian statistical power of $0.873$. The value $RB(0 \,|\,x) = 1.757$ provides evidence in favor of $H_0$. The strength of this evidence is $Str(0\,|\,x)=0.18$, so there is moderate evidence in favor, but certainly not worth claiming that $H_0$ is true. With the background so much greater than the signal, it is hard to discern the signal, at least with such a small sample size. The resulting plausible interval is $(0, 1.87)$, and it contains $92.3\%$ of the posterior probability. It is notable that the plausible interval contains the true value of $\lambda$ even when there is considerable noise. $\blacksquare$ \smallskip

\begin{table}[t]
\centering
\begin{tabular}{lccc}
n & Frequentist Conf. & Bayesian Conf.  & $H_0 :\lambda=0$  \\
\hline
1 & 0.679 & 0.792 & 0.241   \\
5 & 0.818 & 0.884 & 0.109   \\
10 & 0.860 & 0.912 & 0.088   \\
25 & 0.901 & 0.940 & 0.062   \\
50 & 0.922 & 0.954 & 0.046   \\
100 & 0.934 & 0.962 & 0.039   \\
\hline
\end{tabular}
\caption{Frequentist confidence and Bayesian confidence for the plausible interval, and bias against for $H_0$, for several values of $n$ in Example 3.5.}
\label{tab:bias_against_vb}
\end{table}

\begin{table}[t]
\centering
\begin{tabular}{lcc}
\multicolumn{3}{c}{$\delta = 1$} \\
\hline
$n$ & $H_0$ & Bif \\
\hline
1 & 0.606 & 0.809 \\
5 & 0.619 & 0.838 \\
10 & 0.534 & 0.763 \\
25 & 0.301 & 0.576 \\
50 & 0.106 & 0.382 \\
100 & 0.011 & 0.176 \\
\hline
\end{tabular}
\hspace {.5cm}
\begin{tabular}{lcc}

\multicolumn{3}{c}{$\delta = 2$} \\
\hline
$n$ & $H_0$ & Bif \\
\hline
1 & 0.450 & 0.686 \\
5 & 0.283 & 0.536 \\
10 & 0.127 & 0.373 \\
25 & 0.007 & 0.147 \\
50 & 0.000 & 0.043 \\
100 & 0.000 & 0.007 \\
\hline
\end{tabular}
\hspace{.5cm}
\begin{tabular}{lcc}

\multicolumn{3}{c}{$\delta = 3$} \\
\hline
$n$ & $H_0$ & Bif \\
\hline
1 & 0.313 & 0.555 \\
5 & 0.085 & 0.314 \\
10 & 0.012 & 0.155 \\
25 & 0.000 & 0.031 \\
50 & 0.000 & 0.004 \\
100 & 0.000 & 0.000 \\
\hline
\end{tabular}
\caption{Bias in favor for $H_0$ and the plausible interval (Bif) for several values of $n$ and  $\delta$ in Example 3.5.} 
\label{tab:bias_in_favor_split_vb}
\end{table}

\begin{figure}[t]
    \centering
    \includegraphics[width=\linewidth]{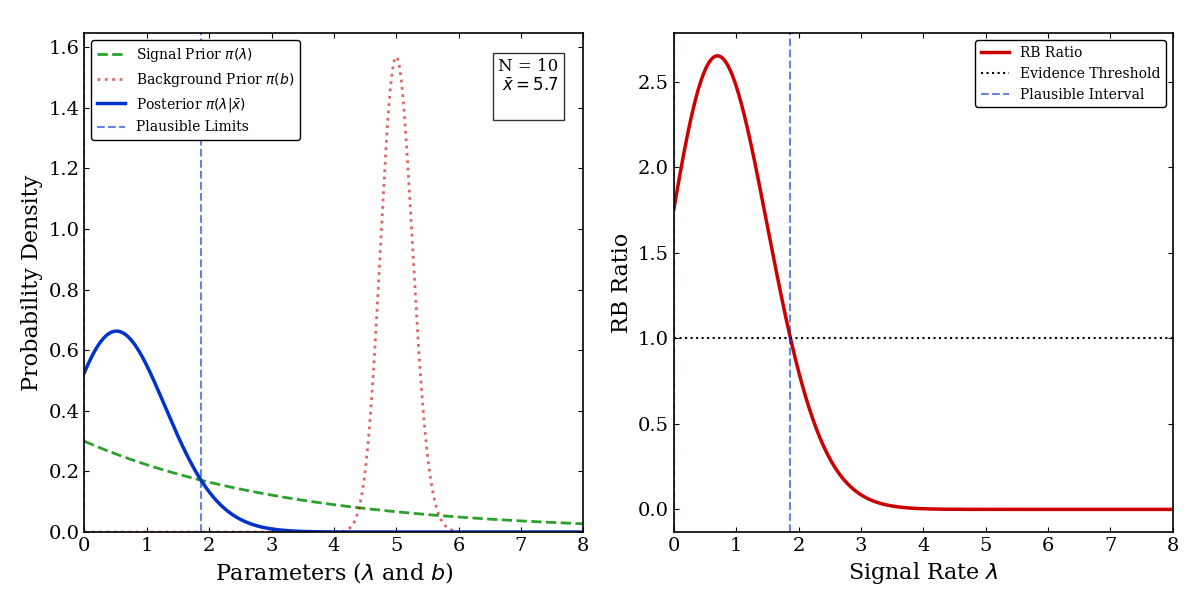}
    \caption{Plots for Example 3.5 based on a generated sample of $n=10$. Left panel: prior on $\lambda$, prior on $b$, posterior density, and plausible interval for $\lambda$. Right panel: Relative belief ratio for $\lambda$ where the horizontal dashed line at 1 marks the evidence cutoff and plausible interval.}
    \label{fig:example35}
\end{figure} 

\noindent\textbf{Example 3.6.} \textit{Karmen II neutrino oscillation experiment.}

In this example, we consider the Karmen II neutrino oscillation experiment: 0 events were observed with expected background $2.88\pm 0.13$ for the first experiment, see Eitel and Zeitnitz (1999), and 11 events were observed with expected background $12.3\pm 0.6$ for the second experiment, see Eitel (2001). To evaluate the signal parameter \(\lambda\), a sequential Bayesian strategy is employed. An elicited initial  $\text{gamma}_{rate}(1,\tau)$ prior for $\lambda$ is updated using the data from experiment 1 to yield a first-round posterior. This posterior then serves as the prior for analyzing the data from experiment 2. The prior on $b$ is elicited based on the interval $(2.75, 3.01)$ for the first experiment and $(2.925, 3.225)$ for the second experiment, in each case the interval contains $95\%$ of the prior probability. The prior on the background $b$ is given by $\text{gamma}_{rate}(1888.38,655.34)$ for the first experiment and $\text{gamma}_{rate}(1617.39,525.66)$ for the second experiment.  .

Table \ref{tab:prior_sensitivity} illustrates the sensitivity of the plausible interval to this initial elicitation, calibrated by specifying target beliefs \(\{0.75, 0.80, 0.90, 0.95, 0.99\}\) over defined intervals \([0, U]\) for \(U \in \{5, 10, 15, 20\}\). It is to be emphasized that the lower limit for each of the plausible intervals computed was always 0. 

For the first experiment (\(t=0\) observed events, \(n=1\)), the resulting upper limits (\(UL_1\)) of the plausible interval exhibit considerable variation, reflecting a heavy dependence on the initial prior configuration. However, as information accumulates in the second experiment (\(t=11\), \(n=4\)), the data overwhelm the prior, causing the sequentially updated upper limits (\(UL_2\)) to robustly stabilize within narrow bounds of roughly 0.5 to 0.7. Furthermore, the analysis consistently indicates a large strength of evidence in favor of the null signal, \(\lambda = 0\), confirming that the evidence for an absence of signal remains decisive regardless of early prior assumptions.

In this example, we refrain from a direct comparison between plausible intervals and Feldman-Cousins confidence intervals (FCCI) due to fundamental structural differences between the two methodologies. First, while FCCI is constructed under the assumption of a fixed, known background \(b\), plausible intervals incorporate systematic variations in \(b\) by treating it as a nuisance parameter with a prior distribution and marginalizing over the joint distribution. Second, within this example, the posterior distribution of the signal parameter, \(\lambda\), from an initial experiment serves as the prior for subsequent analyses, ensuring that plausible intervals condition on accumulated evidence. In contrast, the frequentist formulation of FCCI generates strictly independent intervals for each experiment without a formal mechanism for sequential knowledge integration. Consequently, evaluating these intervals side-by-side would obscure their distinct operational assumptions and yield fundamentally mismatched comparisons. $\blacksquare$

\begin{table}[t]
\centering
\resizebox{\textwidth}{!}{%
\begin{tabular}{ccccccc}
\toprule
\multicolumn{7}{l}{\textit{Belief $= 0.75$}} \\
\midrule
\textbf{UL}\textsubscript{init} & \textbf{UL}\textsubscript{1} & \textbf{Cont}\textsubscript{1} & \textbf{UL}\textsubscript{2} & \textbf{Cont}\textsubscript{2} & $\textbf{RB}_\Lambda (0 \,|\, x)$ & $\textbf{Str}_\Lambda(0 \,|\, x)$ \\
\midrule
5.0 & 1.53 & 0.86 & 0.65 & 0.78 & 1.31 & 0.997 \\
10.0 & 2.11 & 0.91 & 0.71 & 0.79 & 1.33 & 0.996 \\
15.0 & 2.47 & 0.93 & 0.73 & 0.80 & 1.34 & 0.996 \\
20.0 & 2.74 & 0.95 & 0.73 & 0.80 & 1.34 & 0.996 \\
\midrule
\multicolumn{7}{l}{\textit{Belief $= 0.80$}} \\
\midrule
\textbf{UL}\textsubscript{init} & \textbf{UL}\textsubscript{1} & \textbf{Cont}\textsubscript{1} & \textbf{UL}\textsubscript{2} & \textbf{Cont}\textsubscript{2} & $\textbf{RB}_\Lambda (0 \,|\, x)$ & $\textbf{Str}_\Lambda(0 \,|\, x)$ \\
\midrule
5.0 & 1.41 & 0.85 & 0.64 & 0.78 & 1.30 & 0.997 \\
10.0 & 1.98 & 0.90 & 0.70 & 0.79 & 1.32 & 0.996 \\
15.0 & 2.33 & 0.93 & 0.72 & 0.80 & 1.33 & 0.996 \\
20.0 & 2.60 & 0.94 & 0.73 & 0.80 & 1.34 & 0.996 \\
\midrule
\multicolumn{7}{l}{\textit{Belief $= 0.90$}} \\
\midrule
\textbf{UL}\textsubscript{init} & \textbf{UL}\textsubscript{1} & \textbf{Cont}\textsubscript{1} & \textbf{UL}\textsubscript{2} & \textbf{Cont}\textsubscript{2} & $\textbf{RB}_\Lambda (0 \,|\, x)$ & $\textbf{Str}_\Lambda(0 \,|\,x)$ \\
\midrule
5.0 & 1.15 & 0.82 & 0.60 & 0.77 & 1.28 & 0.997 \\
10.0 & 1.68 & 0.87 & 0.67 & 0.79 & 1.31 & 0.997 \\
15.0 & 2.02 & 0.90 & 0.70 & 0.79 & 1.33 & 0.996 \\
20.0 & 2.27 & 0.92 & 0.72 & 0.80 & 1.33 & 0.996 \\
\midrule
\multicolumn{7}{l}{\textit{Belief $= 0.95$}} \\
\midrule
\textbf{UL}\textsubscript{init} & \textbf{UL}\textsubscript{1} & \textbf{Cont}\textsubscript{1} & \textbf{UL}\textsubscript{2} & \textbf{Cont}\textsubscript{2} & $\textbf{RB}_\Lambda (0 \,|\, x)$ & $\textbf{Str}_\Lambda(0 \,|\, x)$ \\
\midrule
5.0 & 0.98 & 0.79 & 0.56 & 0.77 & 1.27 & 0.997 \\
10.0 & 1.47 & 0.85 & 0.65 & 0.78 & 1.30 & 0.997 \\
15.0 & 1.79 & 0.88 & 0.68 & 0.79 & 1.32 & 0.996 \\
20.0 & 2.04 & 0.90 & 0.70 & 0.79 & 1.33 & 0.996 \\
\midrule
\multicolumn{7}{l}{\textit{Belief $= 0.99$}} \\
\midrule
\textbf{UL}\textsubscript{init} & \textbf{UL}\textsubscript{1} & \textbf{Cont}\textsubscript{1} & \textbf{UL}\textsubscript{2} & \textbf{Cont}\textsubscript{2} & $\textbf{RB}_\Lambda (0 \,|\, x)$ & $\textbf{Str}_\Lambda(0 \,|\, x)$ \\
\midrule
5.0 & 0.73 & 0.76 & 0.49 & 0.75 & 1.24 & 0.997 \\
10.0 & 1.15 & 0.82 & 0.60 & 0.77 & 1.28 & 0.997 \\
15.0 & 1.45 & 0.85 & 0.64 & 0.78 & 1.30 & 0.997 \\
20.0 & 1.68 & 0.87 & 0.67 & 0.79 & 1.31 & 0.997 \\
\bottomrule
\end{tabular}%

}
\caption{Sensitivity analysis of the plausible upper limits based on an initial prior elicitation where prior probability \textit{Belief} is allocated to the interval $(0,\textbf{UL}_\text{init})$. Here $\textbf{UL}_\text{i}, \textbf{Cont}_\text{i}$ are, respectively, the upper limits of the plausible interval (lower limit is always 0) and its posterior content based upon experiment i. The evidence, and the strength of the evidence, for the hypothesis $H_0 : \lambda=0$ are presented as $\textbf{RB}_\Lambda (0 \, | \, x)$ and $\textbf{Str}_\Lambda(0 \,|\, x)$, respectively.}

\label{tab:prior_sensitivity}
\end{table}

\section{Code Availability}
The methods described in this paper are implemented in the Python
package \texttt{rbinfer}, freely available at
https://github.com/siqi-zheng/rbinfer.
Main results are reproducible using
the scripts provided in the repository.

\section{Conclusions}

The general issue of making inferences based on a clear definition of what is meant by statistical evidence has been discussed. This leads to what are called relative  belief inferences because they are based on how beliefs change from a priori to a posteriori. This reflects what evidence does, namely, it changes our beliefs. That implies, of course that there are proper prior beliefs that can be used as part of measuring evidence. A reasonable degree of robustness to how these beliefs are expressed is necessary and this is part of the relative belief approach. This robustness depends on the lack of prior-data conflict and checking for this, and suitably modifying the prior when encountered, is also part of the story. 

Given that relative belief inferences are Bayesian in nature, all the nice features associated with that approach, such as being able to quote posterior probabilities for reported intervals, are available. But there is also the necessity to deal with the fundamental question that frequentist inferences address, namely, are the reported inferences reliable? The reliability issue is addressed in relative belief by the bias calculations which are an a priori part of the design of an experiment and fundamentally frequentist in nature. Of course, there may be other a priori calculations that an experimenter chooses to do and control by sample size. There is no contradiction with the actual inferences reported in doing this and, in fact, this could only increase our \textit{confidence} in the reported results. For example, one could control the MSE of the relative belief estimate at certain parameter values if so desired.

Another aspect of the paper has been to compare the interval reported by relative belief with the FCCI. The FCCI clearly does its job as a confidence interval for constrained parameters, but we would argue that there are a number of positive gains to be had by reporting plausible intervals with control over their coverage probabilities. Addressing the evidence issue in a direct way,  makes relative belief a natural approach for statistical reasoning in scientific applications.

\section{References}

\noindent Al-Labadi, L., Alzaatreh, A. and Evans, M. (2025) How to measure
evidence and its strength: Bayes factors or relative belief ratios? Canadian Journal of Statistics, 53: e70015. \newline https://doi.org/10.1002/cjs.70015 \smallskip

\noindent Al-Labadi, L. and Evans, M. (2017) Optimal robustness results for
some Bayesian procedures and the relationship to prior-data conflict. Bayesian
Analysis 12, 3, 702-728.\smallskip

\noindent Biller, S. D. and Oser, S. M. (2015)\ Another look at confidence
intervals: proposal for a more relevant and transparent approach. Nuclear
instruments \& methods in physics research. Section A, Accelerators,
spectrometers, detectors and associated equipment, 2015-02, Vol.774, p.103-119
\smallskip

\noindent Birnbaum, A. (1977) The Neyman-Pearson theory as decision theory,
and as inference theory; with a criticism of the Lindley-Savage argument for
Bayesian theory. Synthese 36, 19-49.\smallskip

\noindent Conrad, J., Botner, O., Hallgren, A., \& P\'{e}rez de los Heros, C.
(2003). Including systematic uncertainties in confidence interval construction
for Poisson statistics. Physical Review. D, Particles and Fields, 67(1),
Article 012002. https://doi.org/10.1103/PhysRevD.67.012002\smallskip

\noindent Cousins, R. D. (1995) Why isn't every physicist a Bayesian?
American Journal of Physics, 63(5), 398--410.
https://doi.org/10.1119/1.17901\smallskip

\noindent Cousins, R. D. and Highland, V. L. (1992) Incorporating systematic
uncertainties into an upper limit. Nuclear Instruments \& Methods in Physics
Research. Section A, Accelerators, Spectrometers, Detectors and Associated
Equipment, 320(1), 331--335.
https://doi.org/10.1016/0168-9002(92)90794-5\smallskip

\noindent Eitel, K., and Zeitnitz, B. (1999). The search for neutrino oscillations → with KARMEN. Nuclear Physics B - Proceedings Supplements, 77(1–3), 212–219. https://doi.org/10.1016/s0920-5632(99)00420-x\smallskip

\noindent Eitel, K. (2001). Latest results of the KARMEN2 experiment. Nuclear Physics B - Proceedings Supplements, 91(1–3), 191–197. https://doi.org/10.1016/s0920-5632(00)00940-3\smallskip

\noindent Englert, B-G. (2026) Lectures on Quantum State Estimation. World Scientific. \smallskip

\noindent Evans (2015) The Measurement of Statistical Evidence Using Relative
Belief. Monographs on Statistics and Applied Probability 144, CRC Press,
Taylor \& Francis Group.\smallskip

\noindent Evans, M. and Guo, Y. (2021) Measuring and controlling bias for some
Bayesian inferences and the relation to frequentist criteria. Entropy 2021,
23(2), 190 doi:10.3390/e23020190.\smallskip

\noindent Evans, M. (2024) The concept of statistical evidence: historical
roots and current developments. Encyclopedia 2024, 4(3), 1201-1216.\smallskip

\noindent Evans, M. and Jang, G-H. (2011a) A limit result for the prior
predictive applied to checking for prior-data conflict. Statistics and
Probability Letters, 81, 1034-1038.\smallskip

\noindent Evans, M. and Jang, G-H. (2011b) Weak informativity and the
information in one prior relative to another. Statistical Science, Vol. 26,
No. 3, 423-439. \smallskip

\noindent Evans, M. Liu, M. Moon, M., Sixta, S. Wei, S. and Yang, S. (2024) On
some problems of Bayesian region construction with guaranteed coverages.
Statistical Papers, 65, 309-334. \newline
doi.org/10.1007/s00362-023-01394-4.\smallskip

\noindent Evans, M. and Moshonov, H. (2006) Checking for prior-data conflict.
Bayesian Analysis, Volume 1, Number 4, 893-914.\smallskip

\noindent Feldman, G. J. and Cousins, R. D. (1998) Unified approach to the
classical analysis of small signals. Physical Review D, 57, 7,
3873-3889.\smallskip

\noindent Gu, Y., Li, W. Evans, M. and Englert, B-G. (2019) Very strong
evidence in favor of quantum mechanics and against local hidden variables from
a Bayesian analysis. Physical Review A 99, 022112 (1-17) (2019) DOI:
10.1103/PhysRevA.99.022112.\smallskip

\noindent Lu, H., Jin, H., Li, Y. and Wang, Z. (2023) Confidence intervals
for a Poisson parameter with background. Communications in Statistics. Theory
and Methods, 52(19), 6794--6805. \newline
https://doi.org/10.1080/03610926.2022.2033268\smallskip

\noindent Lyons, L. (1986) Statistics for Nuclear and Particle Physicists. Cambridge University Press. https://doi.org/10.1017/CBO9781139167710 \smallskip

\noindent Mandelkern, M. (2002) Setting Confidence Intervals for Bounded
Parameters. Statistical Science, 17(2), 149--159.
https://doi.org/10.1214/ss/1030550859\smallskip

\noindent Neyman, J. (1937) Outline of a Theory of Statistical Estimation
Based on the Classical Theory of Probability. Philosophical Transactions of
the Royal Society of London, Series A: Mathematical and Physical Sciences,
236(767), 333--380. https://doi.org/10.1098/rsta.1937.0005\smallskip

\noindent Nott,D., Wang, X., Evans, M., and Englert, B-G. (2020) Checking for
prior-data conflict using prior to posterior divergences. Statistical Science,
35, 2, 234-253.\smallskip

\noindent Oser, S. Fun with confidence intervals. Online notes for Physics 509 at
U. of British Columbia. \newline 
https://phas.ubc.ca/\symbol{126}oser/p509/Lec\_16.pdf \smallskip

\noindent Plante, A. (2020) A Gaussian alternative to using improper confidence intervals. Canadian Journal of Statistics, 48, 4, 773–801. \smallskip

\noindent Plante, A. (2026) Non-negative Gaussian estimation of variance components in random effects models. e70051. https://doi.org/10.1002/cjs.70051 \smallskip

\noindent Popper, K. (1968) The Logic of Scientific Discovery, Harper
Torchbooks. \smallskip

\noindent Roe, B. P. and Woodroofe, M. B. (1998) Improved Probability Method
for Estimating Signal in the Presence of Background.
\newline
https://doi.org/10.48550/arxiv.physics/9812036\smallskip

\noindent Roe, B. P. and Woodroofe, M. B. (2001) Setting confidence belts.
Physical Review. D, Particles and Fields, 63(1), Article 013009.
\newline
https://doi.org/10.1103/PhysRevD.63.013009\smallskip

\noindent Royall, Richard M. (1997) Statistical Evidence: A likelihood
paradigm. London, UK: Chapman \& Hall. \smallskip

\noindent Salmon,W. (1973) Confirmation. Scientific American, 228, 75-81. \smallskip

\noindent Shang, J., Ng, H. K., Sehrawat, A., Li, X. and Englert, B.-G. (2013)
Optimal error regions for quantum state estimation, New J. Phys. 15,
123026.\smallskip

\noindent Teo, Y.S., Shringarpure, S.U., Jeong, H., Prasannan, N., Brecht, B.,
Silberhorn, C., Evans, M., Mogilevtsev, D. and Sanchez-Soto, L.L. (2024a)
Relative-belief inference in quantum information theory. Physical Review A,
July, 2024. DOI: 10.1103/PhysRevA.110.012231.\smallskip

\noindent Teo, Y.S., Jeong, H., Prasannan, N., Brecht, B., Silberhorn, C.,
Evans, M., Mogilevtsev, D. and Sanchez-Soto, L.L. (2024b) Evidence-based
certification of quantum dimensions. Physical Review Letters 133, 050204, DOI:
10.1103/PhysRevLett.133.050204.\smallskip

\end{document}